\documentclass[12pt]{article}

\usepackage[margin=1in]{geometry}

\usepackage{setspace}
\usepackage{amsmath, amssymb, graphicx, natbib}
\usepackage{eurosym}
\usepackage{bbm}
\usepackage[figuresright]{rotating}
\usepackage{algorithm}
\usepackage{algpseudocode}
\usepackage{xr-hyper} 
\usepackage[colorlinks=true, citecolor=blue]{hyperref}
\newtheorem{theorem}{Theorem}

\usepackage{booktabs}
\usepackage{threeparttable}
\usepackage[nomarkers,nolists]{endfloat} %

\begin{document}

\title{\large
\textbf{Why Is the Double-Robust Estimator for Causal Inference }\\
\textbf{Not Doubly Robust for Variance Estimation?}
}
\begin{singlespacing}
{\fontsize{4}{5}\selectfont
\author{
Wu, H.$^{1\ast}$, Shao, L.$^{2\ast}$, Gui, T.T.$^{3}$, Wu, T.-C.$^{2}$, Huang, Z.$^{4}$, Tu, S.$^{2}$,\\[-4pt]
\textbf{Tu, X.M.$^{2}$, Liu, J.$^{1\ast\ast}$, and Lin, T.$^{3}$}\\[-4pt]
$^{1}$Department of Biostatistics, \\[-4pt]
Vanderbilt University Medical Center, Nashville, TN 37232\\[-4pt]
$^{2}$Division of Biostatistics and Bioinformatics\\
UCSD Herbert Wertheim School of Public Health and Human Longevity Science\\[-4pt]
La Jolla, CA 92093\\[-4pt]
$^{3}$Department of Biostatistics, 
University of Florida, Gainesville, FL 32611\\[-4pt]
$^{4}$Department of Statistics, 
University of Florida, Gainesville, FL 32611\\[-4pt]
$^{\ast}$These authors contributed equally to this work.\\[-4pt]
$^{\ast\ast}$Corresponding author: Jinyuan Liu, \texttt{jinyuan.liu@vumc.org}
}
}
\end{singlespacing}

\date{}

\maketitle

\section*{Abstract}
Doubly robust estimators (DRE) are widely used in causal inference because
they yield consistent estimators of average causal effect when at least one
of the nuisance models, the propensity for treatment (exposure) or the outcome regression, is correct. However, double-robustness does not extend to variance
estimation; the influence-function (IF)-based variance estimator is
consistent only when both nuisance parameters are correct. This raises
concerns about applying DRE in practice, where model misspecification is
inevitable. The recent paper by \citealt{shook2025double} (\textit{Biometrics}, 2025, \textbf{81}(2), ujaf054) demonstrated through
Monte Carlo simulations that the IF-based variance estimator is biased.
However, the paper's findings are empirical. The key question remains: why
does the variance estimator fail in double robustness, and under what conditions do
alternatives succeed, such as the ones demonstrated in \citealt{shook2025double}.
In this paper, we develop a formal theory to clarify the efficiency properties
of DRE that underlie these empirical findings. We also introduce alternative
strategies, including a mixture-based framework underlying the
sample-splitting and cross-fitting to achieve valid inference with
misspecified nuisance parameters. Our considerations are illustrated with
simulation and real study data.\\
\textbf{KEY WORDS:}
Efficient influence function; 
Efficient score; 
Joint inference; 
Mixture distribution; 
Sampling-splitting and cross-fitting; 
Variational dependence and independence.

\section{Introduction\label{sec1}}

Doubly robust estimators (DREs), or augmented inverse probability weighting
(AIPW) estimators, are widely used in causal inference 
\citep{laan2003unified,tsiatis2006semiparametric,
ding2018causal,kennedy2024semiparametric}. They yield consistent estimators
of the average causal effect when at least one of the nuisance propensity score (PS) or outcome regression
(OR, or g-computation) model is correctly
specified (\citealt{bang2005doubly}; \citealt{kang2007demystifying} ; %
\citealt{funk2011doubly}). However, discussion of this double-robustness
property has focused almost exclusively on point estimation, with much less
attention given to another critical component for valid inference: the variance estimation.
In particular, the commonly used influence-function (IF)-based variance
estimator is not doubly robust, i.e., the variance estimator is consistent
only when both nuisance models are correctly specified. This raises
concerns about inference validity in practice, since DREs are widely
perceived to achieve double robustness for both point and variance
estimation (\citealt{smith2023application}). \ 

Although IF-based variance estimators for DREs are widely implemented in R
packages such as \texttt{AIPW} (\citealt{zhong2021aipw}) and \texttt{tmle} (%
\citealt{gruber2012tmle}), prior work has noted their limitations. For
example, \citealt{munoz2012population} observed that inference based on the
IF of AIPW, or targeted maximum likelihood estimation (TMLE), is valid only
if both nuisance models are correct. Additionally, related work on causal
inference for Mann-Whitney-Wilcoxon rank-sum test and missing data in
longitudinal studies has used joint inference to estimate both the
target and nuisance parameters to achieve valid inference (e.g., \citealt{chen2016rank}; \citealt{chen2024doubly}). Yet, these approaches have not been
connected to a broader theory of consistent variance estimation for DREs in
causal inference. \ 

Recent work by \citealt{shook2025double} provided the first empirical
investigation of this problem. They performed simulation studies to compare
three approaches: (1) IF-based variance estimation, (2) joint inference
(noted as the sandwich estimator in \citealt{shook2025double}), and (3)
nonparametric bootstrap. Their simulation results showed that only the latter two yield
correct coverage when one nuisance model is misspecified, consistent with prior findings (\citealt{lunceford2004stratification,funk2011doubly}). Although these findings reveal important limitations of IF-based variance estimators, there remains a lack of theoretical explanation for their failure of double robustness in variance estimation, as well as clear guidance on how the latter two approaches correct this bias.

In this paper, we address these gaps by developing a formal theory that delineates the conditions under which variance estimators inherit the double robustness property of point estimators. Our framework extends beyond standard doubly robust causal estimands to accommodate a general vector of target parameters, thereby unifying inference for multivariate semiparametric models. The proposed theory provides formal justification for the empirical findings of \citealt{shook2025double} and clarifies the scope of the double robustness property through the semiparametric efficiency theory (i.e., the efficient influence function). We further introduce alternative strategies that yield valid inference under misspecified nuisance models, thereby strengthening the robustness of DREs in causal analyses. \ 

The remainder of the paper is organized as follows. Section~\ref{sec2}
reviews influence functions and DREs in causal inference. Section~\ref{sec3}
examines one popular type of IF-based variance estimators that remain
unbiased under misspecifed nuisance parameters. Section~\ref{sec4} explains
why the IF-based variance estimator fails under misspecified nuisance and justifies the robust behavior of
joint inference and bootstrap, before introducing a mixture-based alternative to expand our toolkit in
Section~\ref{sec4.4}. Section~\ref{sec5} presents results from simulations
and a real study, and we conclude in Section~\ref{sec6}.

\section{Doubly Robust Estimator for Causal Inference\label{sec2}}

Consider a binary treatments variable $x_{i}$ for the $i$th subject $%
\left( 1\leq i\leq n\right) $,
with $x_{i}=0$ if not treated and $x_{i}=1$ if treated. Let $y_{i}^{x}$ denote the two potential
outcomes and $\mathbf{z}_{i}$ a vector covariates for the $i$th subject. We are interested in inferences about the
average causal effect: 
\begin{equation}
\mu =E\left \{ E(y_{i}\mid x_{i}=1,\mathbf{z}_{i})-E(y_{i}\mid x_{i}=0,%
\mathbf{z}_{i})\right \}. \label{eqn001} 
\end{equation}%

Consider three assumptions: i) counterfactual consistency; ii) ignorability; and iii) positivity (%
\citealt{robins1994estimation}; \citealt{cole2009consistency}; %
\citealt{hernan2020causal}; \citealt{naimi2023defining}). Let $E(x_{i}\mid \mathbf{z}%
_{i})=\eta _{i}\left( \mathbf{z}_{i},\boldsymbol{\psi }\right) $ denote
the mean of a parametric model for the propensity score (PS) parameterized
by $\boldsymbol{\psi }$ and $E(y_{i}\mid x_{i},\mathbf{z}_{i})=Q_{i}(x_{i},%
\mathbf{z}_{i},\boldsymbol{\xi })$ the mean of a parametric linear model for
the outcome regression (OR) parameterized by $\boldsymbol{\xi }$, where $%
y_{i}=x_{i}y_{i}^{1}+(1-x_{i})y_{i}^{0}$ is the observed outcome. 

Then the
widely considered efficient influence function (EIF) for the double robust estimator (DRE) of $\mu $ is (\citealt{tsiatis2006semiparametric,van2011targeted,hines2022demystifying}): \ 
\begin{align}
U_{i}(\boldsymbol{\theta })={}& \frac{x_{i}}{\eta _{i}\left( \boldsymbol{%
\psi }\right) }\{y_{i}-Q_{i}(1,\mathbf{z}_{i},\boldsymbol{\xi })\}-\frac{%
1-x_{i}}{1-\eta _{i}\left( \boldsymbol{\psi }\right) }\{y_{i}-Q_{i}(0,%
\mathbf{z}_{i},\boldsymbol{\xi })\}+  \label{eqn10} \\
& \;+\{Q_{i}(1,\mathbf{z}_{i},\boldsymbol{\xi })-Q_{i}(0,\mathbf{z}_{i},%
\boldsymbol{\xi })\}-\mu,  \notag
\end{align}%
where the joint parameter vector $\boldsymbol{\theta }=\left( \mu ,\boldsymbol{\psi }^{\top },%
\boldsymbol{\xi }^{\top }\right) ^{^{\top }}$ is identifiable under the above three assumptions. In what follows, we will refer to an influence
function (IF)\ such as $U_{i}(\boldsymbol{\theta })$ or a score as an
\textit{estimating function} when it is unnecessary to distinguish the two. \ 

One advantage of the DRE\ $\widehat{\mu }$ obtained by solving these $U_{i}$%
-defined estimating equations, $\sum_{i}U_{i}(\mu ,\widehat{\boldsymbol{\psi 
}}, \widehat{\boldsymbol{\xi }})=0$, with the substitution of some consistent $\widehat{\boldsymbol{\psi }}$ and  $%
\widehat{\boldsymbol{\xi }}$, is
its double robustness, i.e., $\widehat{\mu }$ is consistent if at least one
of the PS and OS models is correct. However, this double-robustness only applies to the
point estimator $\widehat{\mu }$, as shown by simulation results
in \citealt{shook2025double}, the variance estimator based on $\sigma
_{\mu }^{2}=Var\{U_{i}(\boldsymbol{\theta })\}$ is biased, if one of the
nuisance parameters $\boldsymbol{\psi }$ and $\boldsymbol{\xi }$ is
misspecified. \ 

For notational brevity, let $Z_{i}=\left \{ y_{i},x_{i},\mathbf{z}%
_{i}\right
\} $ denote the observed data. For simplicity, and without loss of generality, we assume that the outcome regression (OR) model is correctly specified and fixed at its ground truth, $%
\boldsymbol{\xi =\xi }_{0}$, while the propensity score (PS) model may be misspecified. We parameterize this potentially misspecified PS by $\boldsymbol{\psi }^{\prime }$ and focus on its role in what follows (thus no need to estimate $\boldsymbol{\xi}$). We redefine the full parameter vector as $\boldsymbol{\theta }=\left( \mu ,%
\boldsymbol{\psi }^{\top }\right) ^{\top }$, where $\mu$ is the target estimand and $\boldsymbol{\psi }$ is the correctly specified PS parameters. Under a misspecified PS model, the parameter vector becomes $\boldsymbol{\theta }^{\prime }=\left( \mu ,%
\boldsymbol{\psi }^{\prime \top }\right) ^{\top }$, where $%
\boldsymbol{\psi }^{\prime }$ denotes the working (possibly incorrect) PS parameters. 

The efficient influence function (EIF) $U_{i}(\boldsymbol{\theta })$ in (\ref%
{eqn10}) is derived under the correctly specified PS model with parameters $\boldsymbol{\psi }$. In this setting, $U_{i}(%
\boldsymbol{\theta })$ is orthogonal to the nuisance tangent space associated with $%
\boldsymbol{\psi }$, and the asymptotic variance of $\widehat{\mu }$ equals the
variance of $U_{i}(\boldsymbol{\theta })$ under the data generating process (%
\citealt{tsiatis2006semiparametric}). Accordingly, a consistent estimator $\widehat{\boldsymbol{\psi }}$ can be substituted to solve $\sum_{i=1}^{n}U_{i}(\mu,\widehat{\boldsymbol{\psi }}) = 0$ for $\widehat{\mu }$. The variance of $\widehat{\mu }$ is then estimated using the plug-in formula 
\[
\widehat{\sigma }_{\mu }^{2}=\frac{1}{n}\sum_{i=1}^{n}U_{i}^{2}(%
\widehat{\boldsymbol{\theta }}),\text{ }  \widehat{\boldsymbol{\theta }}=( 
\widehat{\mu },\widehat{\boldsymbol{\psi }}^{\top }) ^{\top }.
\]
This so-called plug-in rule provides a consistent variance estimator only when $U_{i}(\boldsymbol{\theta })$ is indeed the EIF (%
\citealt{kennedy2019nonparametric}), which requires that $\boldsymbol{\psi }%
^{\prime }\boldsymbol{=\psi }$. 

We argue that, when the specified PS model is incorrect ($%
\boldsymbol{\psi }^{\prime }\neq\boldsymbol{ \psi }$), the influence function $U_{i}(\boldsymbol{\theta }%
^{\prime })$ is no longer efficient: it does not remain orthogonal to the nuisance tangent space associated with $%
\boldsymbol{\psi }^{\prime }$. Consequently, the IF-based variance estimator via plug-in, $\widehat{%
\sigma }_{\mu }^{2}=(1/n)\sum_{i=1}^{n}U_{i}^{2}(\widehat{\boldsymbol{%
\theta }}^{\prime })$, is inconsistent under PS misspecification, as demonstrated in \citealt{shook2025double}. To address this issue, they proposed two alternative procedures that yield consistent estimators of $Var\left( \widehat{\mu }\right) $ even when the PS model is misspecified.

Before examining why the plug-in variance estimator fails to remain double robustness and how the proposed corrections work, we first introduce the key concept of variational (in)dependence. Throughout, we assume all estimators are regular and asymptotically linear (RAL) unless otherwise specified.

\section{Variational (In)dependence and Joint Inference\label{sec3}}

Many efficient estimating functions remain efficient under misspecified
nuisance parameters. A well-known example is the class of robust efficient estimating functions for restricted moment models \citep{tsiatis2006semiparametric}. For such estimating functions, the plug-in rule continues to yield consistent variance estimation, since the efficiency is preserved regardless of nuisance specification. We provide additional illustrations of this type in the Supplemental Material Web Appendix A.1. \ 

However, this robustness does not extend to the causal DRE\ under a
misspecified PS. Specifically, the EIF for the causal DRE, $U_{i}(\boldsymbol{\theta }^{\prime })=U\left( Z_{i},\mu ,%
\boldsymbol{\psi }^{\prime }\right) $ is efficient only when the specified PS model coincides with the truth, i.e., $\boldsymbol{%
\psi }^{\prime }=\boldsymbol{\psi }$. \ When $\boldsymbol{\psi }%
^{\prime }\neq \boldsymbol{\psi }$, $U\left( Z_{i},\mu ,%
\boldsymbol{\psi }^{\prime }\right) $ is no longer efficient, and its variability alone is not
sufficient to determine the asymptotic variance of the DRE estimator $\widehat{\mu }$.

This inefficiency arises from the phenomenon of \textit{variational dependence} between the parameters of
interest and nuisance parameters. In the next section, we discuss this concept in detail and describe how to account for such dependence in a general semiparametric setting. 

\subsection{Variational Dependence and Joint Inference with Inefficient
Scores\label{sec3.1}}

Let $\boldsymbol{\theta }=\left( \boldsymbol{\mu }^{\top },\boldsymbol{\psi }%
^{\top }\right) ^{\top }$, where $\boldsymbol{\mu }$ is a $p$-dimensional vector of target parameters and $\boldsymbol{\psi }$
is a $q$-dimensional vector of nuisance
parameters. This setup is general and not restricted to the causal DRE setting with a scalar $\mu$. Define the joint (efficient) score for $\boldsymbol{\theta }$ as $%
\boldsymbol{S}_{i}\left( \boldsymbol{\theta }\right) =\left( \boldsymbol{S}%
_{\boldsymbol{\mu} ,i}^{\top }\left( \boldsymbol{\theta }\right) ,\boldsymbol{S}_{ \boldsymbol{\psi}
,i}^{\top }\left( \boldsymbol{\theta }\right) \right) ^{\top }$, where $%
\boldsymbol{S}_{\boldsymbol{\mu} ,i}$\ and $\boldsymbol{S}_{\boldsymbol{\psi} ,i}$ are the
scores for $\boldsymbol{\mu }$ and $\boldsymbol{\psi }$, respectively. \ Similarly, let the joint estimating function be $\boldsymbol{W}%
_{i}\left( \boldsymbol{\theta }\right) =\left( \boldsymbol{U}_{i}^{\top
}\left( \boldsymbol{\theta }\right) ,\boldsymbol{V}_{i}^{\top }\left( 
\boldsymbol{\theta }\right) \right) ^{\top }$, where $\boldsymbol{U}_{i}$\ and 
$\boldsymbol{V}_{i}$ are the influence functions for $\boldsymbol{\mu }$ and $%
\boldsymbol{\psi }$, respectively. We first consider the case when $\boldsymbol{W}_{i}\left( \boldsymbol{%
\theta }\right) =\boldsymbol{S}_{i}\left( \boldsymbol{\theta }\right) $ to
illustrate the notion of variational dependence. \ 

Under mild regularity conditions, the Fisher information for $\boldsymbol{%
\theta }$ is given by $\boldsymbol{I}\left( 
\boldsymbol{\theta }\right) =E\{-\frac{\partial }{\partial \boldsymbol{%
\theta }}\boldsymbol{W}_{i}\left( \boldsymbol{\theta }\right) \}$. This block matrix can be written as:%
\begin{equation*}
\boldsymbol{I}\left( \boldsymbol{\theta }\right) =\left( 
\begin{array}{cc}
\boldsymbol{I}_{11}\left( \boldsymbol{\theta }\right) & \boldsymbol{I}%
_{12}\left( \boldsymbol{\theta }\right) \\ 
\boldsymbol{I}_{12}\left( \boldsymbol{\theta }\right) & \boldsymbol{I}%
_{22}\left( \boldsymbol{\theta }\right)%
\end{array}%
\right) =\left( 
\begin{array}{cc}
Var\{ \boldsymbol{U}_{i}\left( \boldsymbol{\theta }\right) \} & E\{%
\boldsymbol{U}_{i}\left( \boldsymbol{\theta }\right) \boldsymbol{V}%
_{i}^{\top }\left( \boldsymbol{\theta }\right) \} \\ 
E\{ \boldsymbol{V}_{i}\left( \boldsymbol{\theta }\right) \boldsymbol{U}%
_{i}^{\top }\left( \boldsymbol{\theta }\right) \} & Var\{ \boldsymbol{V}%
_{i}\left( \boldsymbol{\theta }\right) \}%
\end{array}%
\right) ,
\end{equation*}%
where the off-diagonal blocks $\boldsymbol{I}_{12}=E\{%
\boldsymbol{U}_{i}\left( \boldsymbol{\theta }\right) \boldsymbol{V}%
_{i}^{\top }\left( \boldsymbol{\theta }\right) \}$ capture the variational dependence between $\boldsymbol{U}_{i}$ and $\boldsymbol{V}_{i}$.
\ Let $\widehat{\boldsymbol{\theta }}=\left( \widehat{\boldsymbol{\mu }}%
^{\top },\widehat{\boldsymbol{\psi }}^{\top }\right) ^{\top }$ be the
estimator from solving the score equations:\ 
\begin{equation}
\sum_{i=1}^{n}\boldsymbol{W}_{i}\left( \boldsymbol{\theta }\right)
=\sum_{i=1}^{n}\left( 
\begin{array}{c}
\boldsymbol{U}_{i}\left( \boldsymbol{\theta }\right) \\ 
\boldsymbol{V}_{i}\left( \boldsymbol{\theta }\right)%
\end{array}%
\right) =\mathbf{0}.  \label{eqn200}
\end{equation}%
By the asymptotic linearity of $\widehat{\boldsymbol{\theta }}$ \citep{hampel1974influence,tsiatis2006semiparametric}, together with the central
limit theorem (CLT) and Slutsky's theorem, we have:%
\begin{equation}
\sqrt{n}\left( \widehat{\boldsymbol{\theta }}-\boldsymbol{\theta }%
_{0}\right) =\frac{\sqrt{n}}{n}\sum_{i=1}^{n}\boldsymbol{I}^{-1}\left( 
\boldsymbol{\theta }_{0}\right) \boldsymbol{W}_{i}\left( \boldsymbol{\theta }%
_{0}\right) +\mathbf{o}_{p}\left( 1\right) \rightarrow _{d}N\left( \mathbf{0}%
,\Sigma _{\boldsymbol{\theta }}\right),  \label{eqn240}
\end{equation}%
where $\rightarrow_{d}$ denotes convergence in distribution and $\boldsymbol{\theta }%
_{0}=\left( \boldsymbol{\mu }_{0}^{\top },\boldsymbol{\psi }_{0}%
^{\top }\right) ^{\top }$ is the ground truth.

Therefore, the asymptotic variance of $\widehat{\boldsymbol{\theta }}$ equals the
inverse of the Fisher: 
\begin{equation}
\Sigma _{\boldsymbol{\theta }}=Var\{ \boldsymbol{I}^{-1}\left( \boldsymbol{%
\theta }_{0}\right) \mathbf{W}_{i}\left( \boldsymbol{\theta }_{0}\right) \}=%
\boldsymbol{I}^{-1}\left( \boldsymbol{\theta }_{0}\right) Var\{ \mathbf{W}%
_{i}\left( \boldsymbol{\theta }_{0}\right) \} \boldsymbol{I}^{-1}\left( 
\boldsymbol{\theta }_{0}\right) =\boldsymbol{I}^{-1}\left( \boldsymbol{%
\theta }_{0}\right) ,  \notag
\end{equation}%
which similarly decomposes into: 
\begin{equation*}
\Sigma _{\boldsymbol{\theta }}=\left( 
\begin{array}{cc}
\Sigma _{\boldsymbol{\mu }} & \Sigma _{\boldsymbol{\mu }\boldsymbol{\psi }}
\\ 
\Sigma _{\boldsymbol{\psi }\boldsymbol{\mu }} & \Sigma _{\boldsymbol{\psi }}%
\end{array}%
\right) =\left( 
\begin{array}{cc}
\boldsymbol{I}_{11} & \boldsymbol{I}_{12} \\ 
\boldsymbol{I}_{21} & \boldsymbol{I}_{22}%
\end{array}%
\right) ^{-1}.
\end{equation*}%
It follows immediately that for the parameter of interest $\boldsymbol{\mu}$: 
\begin{equation}
\sqrt{n}\left( \widehat{\boldsymbol{\mu }}-\boldsymbol{\mu }_{0}\right)
\rightarrow _{d}N\left( \mathbf{0},\Sigma _{\boldsymbol{\mu }}\right) ,\quad
\Sigma _{\boldsymbol{\mu }}=\left( \boldsymbol{I}_{11}-%
\boldsymbol{I}_{12}\boldsymbol{I}_{22}^{-1}\boldsymbol{I}_{12}^{\top
}\right) ^{-1}.  \label{eqn280}
\end{equation}%

Thus, the variance of $\widehat{\boldsymbol{\mu }}$ is not $\boldsymbol{%
I}_{11}^{-1}$ (the inverse information from $\boldsymbol{U}_{i}\left( \boldsymbol{\theta }\right)$ alone), but
rather, the larger matrix $\left( \boldsymbol{I}_{11}-\boldsymbol{I}_{12}%
\boldsymbol{I}_{22}^{-1}\boldsymbol{I}_{12}^{\top }\right) ^{-1}$ ($%
\mathbf{A}\leq \mathbf{B}$ is defined as $\mathbf{B}-\mathbf{A}$ being a
positive semi-definite matrix). \ 

When $\boldsymbol{\mu }$ and $\boldsymbol{\psi }$ are variationally
independent, $\boldsymbol{I}_{12}\left( \boldsymbol{\theta }%
_{0}\right) =E\{ \boldsymbol{\ U}_{i}\left( \boldsymbol{\theta }_{0}\right) 
\boldsymbol{V}_{i}^{\top }\left( \boldsymbol{\theta }_{0}\right) \}=\mathbf{0%
}$, and $\Sigma _{\boldsymbol{\mu }}=\boldsymbol{I}_{11}^{-1}\left( 
\boldsymbol{\theta }_{0}\right) =Var\{ \boldsymbol{\ U}_{i}\left( 
\boldsymbol{\theta }_{0}\right) \}$, and vice versa. \ In this case, (\ref{eqn240}) simplifies to:\ 
\begin{align*}
\sqrt{n}\left[ \left( 
\begin{array}{c}
\widehat{\boldsymbol{\mu }} \\ 
\widehat{\boldsymbol{\psi }}%
\end{array}%
\right) -\left( 
\begin{array}{c}
\boldsymbol{\mu }_{0} \\ 
\boldsymbol{\psi }_{0}%
\end{array}%
\right) \right] & =\frac{\sqrt{n}}{n}\sum_{i=1}^{n}\left( 
\begin{array}{cc}
\boldsymbol{I}_{11}^{-1} & \boldsymbol{0} \\ 
\boldsymbol{0} & \boldsymbol{I}_{22}^{-1}%
\end{array}%
\right) \boldsymbol{W}_{i}\left( \boldsymbol{\theta }_{0}\right) +\mathbf{o}%
_{p}\left( 1\right) \\
& =\frac{\sqrt{n}}{n}\sum_{i=1}^{n}\left( 
\begin{array}{cc}
\boldsymbol{I}_{11}^{-1}\boldsymbol{U}_{i}\left( \boldsymbol{\theta }%
_{0}\right) & \\ \boldsymbol{I}_{22}^{-1}\boldsymbol{V}_{i}\left( \boldsymbol{\theta }%
_{0}\right)%
\end{array}%
\right)  +\mathbf{o}%
_{p}\left( 1\right) \rightarrow _{d}N\left( \mathbf{0},\Sigma _{\boldsymbol{\theta }%
}\right) .
\end{align*}%
Thus, $\boldsymbol{U}_{i}\left( \boldsymbol{\theta }\right) $ is efficient when $\boldsymbol{\mu }$ and $\boldsymbol{\psi }$ are
variationally independent. Plugging-in any $\sqrt{n}$-consistent estimator $\widehat{\boldsymbol{\psi}}$, the $\widehat{\boldsymbol{\mu}}$ obtained from the $\boldsymbol{U}_{i}$%
-based estimating equations
\begin{equation}
\sum_{i=1}^{n}\boldsymbol{U}_{i}( \boldsymbol{\mu },\widehat{%
\boldsymbol{\psi }}) =\sum_{i=1}^{n}\boldsymbol{U}_{i}( Z_{i},%
\boldsymbol{\mu },\widehat{\boldsymbol{\psi }}) =\mathbf{0}
\label{eqn300}
\end{equation}%
 is consistent and asymptotically normal. More importantly, its asymptotic variance
is the inverse of the Fisher $\boldsymbol{I}_{11}\left( 
\boldsymbol{\theta }_{0}\right) $ (for $\boldsymbol{U}_{i}\left( \boldsymbol{%
\theta }\right) $ alone), with a consistent estimator given by: 
\begin{equation}
\widehat{\Sigma }_{\boldsymbol{\mu }}=\boldsymbol{I}_{11}^{-1}( 
\widehat{\boldsymbol{\mu }},\widehat{\boldsymbol{\psi }}) =\left \{ 
\frac{1}{n}\sum_{i=1}^{n}\boldsymbol{U}_{i}( \widehat{\boldsymbol{\mu }}%
,\widehat{\boldsymbol{\psi }}) \boldsymbol{U}_{i}^{\top }( 
\widehat{\boldsymbol{\mu }},\widehat{\boldsymbol{\psi }}) \right \}
^{-1}.  \label{eqn310}
\end{equation}

But if $\boldsymbol{\mu }$ and $\boldsymbol{\psi }$ are not variationally
independent, then $\boldsymbol{I}_{12}\left( \boldsymbol{\theta }_{0}\right) \neq 
\mathbf{0}$ and the variance of target parameter is $\Sigma _{\boldsymbol{\mu }}=\left( \boldsymbol{I}_{11}-%
\boldsymbol{I}_{12}\boldsymbol{I}_{22}^{-1}\boldsymbol{I}_{12}^{\top
}\right) ^{-1}$, which is greater than $\boldsymbol{I}_{11}^{-1}\left( \boldsymbol{\theta }%
_{0}\right) $. Further, $\boldsymbol{U}_{i}\left( \boldsymbol{\theta }%
\right) $ is inefficient, since its variability alone does not capture the uncertainty in $\widehat{\boldsymbol{\mu }}$. \ One natural remedy is to
jointly estimate $\boldsymbol{\mu }$ and $\boldsymbol{\psi }$\ using the full estimating equations $%
\boldsymbol{W}_{i}\left( \boldsymbol{\theta }\right) $ in (\ref{eqn200}), which incorporate the covariation between $%
\boldsymbol{U}_{i}\left( \boldsymbol{\theta }\right) $ and $\boldsymbol{V}%
_{i}\left( \boldsymbol{\theta }\right) $. This joint estimation continues to yield the correct asymptotic variance for $\widehat{\boldsymbol{\mu }}$, as shown by \eqref{eqn240} and \eqref{eqn280}. \ 

\subsection{Inference with Efficient Score
Function\label{sec3.2}}

In addition to joint inference, one can construct an efficient version of the score $%
\boldsymbol{U}_{i}\left( \boldsymbol{\theta }\right) $ that accounts for the
covariation of $\boldsymbol{U}_{i}\left( \boldsymbol{\theta }\right) $ and $%
\boldsymbol{V}_{i}\left( \boldsymbol{\theta }\right) $. Define the efficient score: 
\begin{equation}
\boldsymbol{U}_{i}^{eff}\left( \boldsymbol{\theta }\right) =\boldsymbol{U}%
_{i}\left( \boldsymbol{\theta }\right) -\boldsymbol{I}_{12}\left( \boldsymbol{\theta }\right) \boldsymbol{I}%
_{22}^{-1}\left( \boldsymbol{\theta }\right) \boldsymbol{V}_{i}\left( 
\boldsymbol{\theta }\right).  \label{eqn320}
\end{equation}%

By construction, $\boldsymbol{U}_{i}^{eff}\left( 
\boldsymbol{\theta }\right) $ is uncorrelated with $\boldsymbol{V}_{i}\left( \boldsymbol{%
\theta }\right) $ and is efficient for estimating $\boldsymbol{\mu }$ (see proof in Supplementary Material Web Appendix A.2). The second term $%
\boldsymbol{I}_{12}\left( \boldsymbol{\theta }\right) \boldsymbol{I}%
_{22}^{-1}\left( \boldsymbol{\theta }\right) \boldsymbol{V}_{i}\left( 
\boldsymbol{\theta }\right) $ in \eqref{eqn320} can be viewed as the projection of $\boldsymbol{U}_{i}\left( 
\boldsymbol{\theta }\right) $ onto the nuisance tangent space spanned by the score $\boldsymbol{V}_{i}\left( 
\boldsymbol{\theta }\right) $ (\citealt{tsiatis2006semiparametric}). It is straightforward that the variance of the efficient score is: 
\begin{equation}
\boldsymbol{I}_{11}^{eff}\left( \boldsymbol{\theta }_{0}\right) =Var\{ 
\boldsymbol{U}_{i}^{eff}\left( \boldsymbol{\theta }_{0}\right)  \}
=Var\left \{ \boldsymbol{U}_{i}\left( \boldsymbol{\theta }_{0}\right) -%
\boldsymbol{I}_{12}\boldsymbol{I}_{22}^{-1}\boldsymbol{V}_{i}\left( 
\boldsymbol{\theta }_{0}\right) \right \} =\boldsymbol{I}_{11}-\boldsymbol{I}%
_{12}\boldsymbol{I}_{22}^{-1}\boldsymbol{I}_{12}^{\top }.  \notag
\end{equation}%
Replacing $\boldsymbol{U}_{i}\left( \boldsymbol{\theta }%
\right) $ with $\boldsymbol{U}_{i}^{eff}\left( \boldsymbol{\theta }\right) $ in the estimating equations (\ref{eqn300}) yields an estimator $%
\widehat{\boldsymbol{\mu }}^{eff}$ that is consistent and asymptotically normal, with
variance $\Sigma
_{\boldsymbol{\mu }}^{eff}=\boldsymbol{I}_{11}^{eff}=\left( \boldsymbol{I}%
_{11}-\boldsymbol{I}_{12}\boldsymbol{I}_{22}^{-1}\boldsymbol{I}_{12}^{\top
}\right) ^{-1}$, exactly the same as the asymptotic variance $\Sigma _{%
\boldsymbol{\boldsymbol{\mu }}}$ obtained
from the joint inference in (\ref{eqn280}). \ 

\subsection{Joint Inference with Inefficient Influence Function\label{sec3.3}%
}

Now suppose that $\boldsymbol{U}_{i}\left( \boldsymbol{\theta }\right) $, corresponding to the target parameter $\boldsymbol{\mu}$, is
the influence function (IF), a normalized version of the score, which might be inefficient, and that $\boldsymbol{V}%
_{i}\left( \boldsymbol{\theta }\right) $ is either an IF or a score. Although $%
\boldsymbol{W}_{i}\left( \boldsymbol{\theta }\right) $ is a general estimating
function (not necessarily a score), we may still define a \textquotedblleft Fisher information" by: 
\begin{equation}
\boldsymbol{I}\left( \boldsymbol{\theta }\right) =E\left \{ -\frac{\partial 
}{\partial \boldsymbol{\theta }}\boldsymbol{W}_{i}\left( \boldsymbol{\theta }%
\right) \right \} =E\left \{ \boldsymbol{W}_{i}\left( \boldsymbol{\theta }%
\right) \boldsymbol{S}_{i}^{\top }\left( \boldsymbol{\theta }\right) \right
\} \neq E\left \{ \boldsymbol{W}_{i}\left( \boldsymbol{\theta }\right) 
\boldsymbol{W}_{i}^{\top }\left( \boldsymbol{\theta }\right) \right \}. 
\notag
\end{equation}%
Namely, it is not equal to the variance of $%
\boldsymbol{W}_{i}\left( \boldsymbol{\theta }\right) $ in general. \ 

Let $\widehat{\boldsymbol{\theta }}$ be the estimator from the joint inference using $%
\boldsymbol{W}_{i}\left( \boldsymbol{\theta }\right) $ as in (\ref{eqn200}). The joint influence function can be derived as $\boldsymbol{\varphi }_{i}\left( \boldsymbol{\theta }\right) =%
\boldsymbol{I}^{-1}\left( \boldsymbol{\theta }\right) \boldsymbol{W}%
_{i}\left( \boldsymbol{\theta }\right) $, with 
\begin{equation*}
\sqrt{n}\left( \widehat{\boldsymbol{\theta }}-\boldsymbol{\theta }%
_{0}\right) =\frac{\sqrt{n}}{n}\sum_{i=1}^{n}\boldsymbol{\varphi }_{i}\left( 
\boldsymbol{\theta }_{0}\right) +\mathbf{o}_{p}\left( 1\right) \rightarrow
_{d}N\left( \mathbf{0},\Sigma _{\boldsymbol{\theta }}\right).
\end{equation*}%
The asymptotic variance $\Sigma _{\boldsymbol{\theta }}$ is the variance of the IF under $\boldsymbol{\theta }_{0}$ (\citealt{kennedy2024semiparametric}): 
\begin{equation*}
\Sigma _{\boldsymbol{\theta }}=Var\{ \boldsymbol{\varphi }_{i}\left( 
\boldsymbol{\theta }_{0}\right) \} =\boldsymbol{I}^{-1}\left( 
\boldsymbol{\theta }_{0}\right) E\left \{ \boldsymbol{W}_{i}\left( 
\boldsymbol{\theta }_{0}\right) \boldsymbol{W}_{i}^{\top }\left( 
\boldsymbol{\theta }_{0}\right) \right \} \boldsymbol{I}^{-1}\left( 
\boldsymbol{\theta }_{0}\right) \neq \boldsymbol{I}^{-1}\left( \boldsymbol{\theta }_{0}\right), 
\end{equation*}%
which has the sandwich variance form. Unlike the score case, $\Sigma _{\boldsymbol{\theta }}$ is no longer
equal to the inverse of this \textquotedblleft Fisher information" even under the joint inference.

If $\boldsymbol{\mu }$ and $\boldsymbol{\psi }$ are variationally independent \citep{tsiatis2006semiparametric},
then 
\begin{equation*}
E\left \{ \boldsymbol{U}_{i}\left( \boldsymbol{\theta }\right) \boldsymbol{S}%
_{\boldsymbol{\mu },i}^{\top }(\boldsymbol{\theta })\right \} =\mathbf{I}%
_{p},\mathbf{\quad }E\left \{ \boldsymbol{U}_{i}\left( \boldsymbol{\theta }%
\right) \boldsymbol{S}_{\boldsymbol{\psi },i}^{\top }\left( \boldsymbol{%
\theta }\right) \right \} =\boldsymbol{0}, \quad   E\left\{ \boldsymbol{V}_{i}\left( \boldsymbol{\theta }\right)  \boldsymbol{S}^{\top}_{\boldsymbol{\mu },i}\left( \boldsymbol{\theta }%
\right) \right\} = \mathbf{0}
\end{equation*}%
where $p=dim(\boldsymbol{\mu})$ and $\mathbf{I}_{p}$
is the $p\times p$ identity matrix. In this case, the \textquotedblleft Fisher information"
has a block-diagonal form: 
\begin{equation*}
\boldsymbol{I}\left( \boldsymbol{\theta }\right) =\left( 
\begin{array}{cc}
E\left \{ \boldsymbol{U}_{i}\left( \boldsymbol{\theta }\right) \boldsymbol{S}%
_{\boldsymbol{\mu },i}^{\top }\right \} & E\left \{ \boldsymbol{U}%
_{i}\left( \boldsymbol{\theta }\right) \boldsymbol{S}_{\boldsymbol{\psi }%
,i}^{\top }\left( \boldsymbol{\theta }\right) \right \} \\ 
E\{ \boldsymbol{V}_{i}\left( \boldsymbol{\theta }\right)  \boldsymbol{S}^{\top}_{\boldsymbol{\mu },i}\left( \boldsymbol{\theta }%
\right) \}
& E\left \{ \boldsymbol{V}_{i}\left( \boldsymbol{\theta }\right) \boldsymbol{%
S}_{\boldsymbol{\psi },i}^{\top }\left( \boldsymbol{\theta }\right) \right \}%
\end{array}%
\right) =\left( 
\begin{array}{cc}
\mathbf{I}_{p} & \boldsymbol{0} \\ 
\boldsymbol{0} & E\left\{\boldsymbol{\ V}_{i}\left( \boldsymbol{\theta }%
\right) \boldsymbol{S}_{\boldsymbol{\psi },i}^{\top }\left( \boldsymbol{%
\theta }\right) \right \}%
\end{array}%
\right).
\end{equation*}%

Substituting into \eqref{eqn240}, we obtain:%
\begin{align*}
\sqrt{n}\left \{ \left( 
\begin{array}{c}
\widehat{\boldsymbol{\mu }} \\ 
\widehat{\boldsymbol{\psi }}%
\end{array}%
\right) -\left( 
\begin{array}{c}
\boldsymbol{\mu }_{0} \\ 
\boldsymbol{\psi }_{0}%
\end{array}%
\right) \right \} & =\frac{\sqrt{n}}{n}\sum_{i=1}^{n}\left( 
\begin{array}{cc}
\mathbf{I}_{p} & \boldsymbol{0} \\ 
\boldsymbol{0} & E\left \{ \boldsymbol{V}_{i}\left( \boldsymbol{\theta }%
_{0}\right) \boldsymbol{S}_{\boldsymbol{\psi },i}^{\top }\left( \boldsymbol{%
\theta }_{0}\right) \right \}%
\end{array}%
\right) ^{-1}\left( 
\begin{array}{c}
\boldsymbol{U}_{i}\left( \boldsymbol{\theta }_{0}\right) \\ 
\boldsymbol{V}_{i}\left( \boldsymbol{\theta }_{0}\right)%
\end{array}%
\right) +\mathbf{o}_{p}\left( 1\right) \\
& =\frac{\sqrt{n}}{n}\sum_{i=1}^{n}\left( 
\begin{array}{c}
\boldsymbol{U}_{i}\left( \boldsymbol{\theta }_{0}\right) \\ 
E^{-1}\left \{ \boldsymbol{V}_{i}\left( \boldsymbol{\theta }_{0}\right) 
\boldsymbol{S}_{\boldsymbol{\psi },i}^{\top }\left( \boldsymbol{\theta }%
_{0}\right) \right \} \boldsymbol{V}_{i}\left( \boldsymbol{\theta }%
_{0}\right)%
\end{array}%
\right) +\mathbf{o}_{p}\left( 1\right).
\end{align*}%
Therefore, when $\boldsymbol{\psi }$ can be consistently estimated, $\boldsymbol{\mu }$ may again be estimated from the $%
\boldsymbol{U}_{i}$-defined estimating equations (\ref{eqn300}), and
its asymptotic variance can be consistently estimated by the plug-in rule using the sample variance of $\boldsymbol{U}%
_{i}\left( \boldsymbol{\theta }\right) $ as in (\ref%
{eqn310}). However, if $\boldsymbol{\mu }$ and $\boldsymbol{\psi }$ are not variationally independent, this plug-in variance estimator is no longer valid.

\section{Joint Inference Under Misspecified Nuisance in Causal Inference\label{sec4}}

\subsection{Sandwich Variance and Joint Inference\label{sec4.1}}

We now return to the problem of doubly robust estimation (DRE) of (\ref{eqn001}) in causal inference. When the nuisance parameter is misspecified, bias in the variance estimation of the DRE $\widehat{\mu }$ arises primarily from variational dependence between $\mu $ and the misspecified $\boldsymbol{\psi }^{\prime }$. In this case, the asymptotic variance of $\widehat{\mu}$ is no longer given by the variance of the scalar influence function (IF) $U_i(\boldsymbol{\theta})$ in \eqref{eqn10}, which was derived under the data generating process where the propensity score (PS) model is correctly specified with parameter $\boldsymbol{\psi}$. To correct for this,  we can still account for the induced covariation as follows. 

Let $%
\boldsymbol{\theta }^{\prime }=\left( \mu ,\boldsymbol{\psi }^{\prime \top
}\right) {^{\top }}$, and denote by $\boldsymbol{V}_{i}\left( \boldsymbol{\theta }^{\prime }\right) $ the score function for the misspecified nuisance $\boldsymbol{\psi }%
^{\prime }$. Define the joint estimating function $\boldsymbol{W}%
_{i}\left( \boldsymbol{\theta }^{\prime }\right) =\left( U_{i}\left( \boldsymbol{\theta }^{\prime }\right) ,\boldsymbol{V}_{i}^{\top }\left( 
\boldsymbol{\theta }^{\prime }\right) \right) ^{\top }$, with the corresponding \textquotedblleft Fisher information": 
\begin{align}
\boldsymbol{I}\left( \boldsymbol{\theta }^{\prime }\right)
&=E\{-\frac{\partial }{\partial \boldsymbol{\theta }^{\prime }}\boldsymbol{W}%
_{i}\left( \boldsymbol{\theta }^{\prime }\right) \}=E\{ \boldsymbol{W}%
_{i}\left( \boldsymbol{\theta }^{\prime }\right) \boldsymbol{S}_{i}^{\top
}\left( \boldsymbol{\theta }^{\prime }\right) \}=\left( 
\begin{array}{cc}
I_{11}\left( \boldsymbol{\theta }^{\prime }\right) & 
\boldsymbol{I}_{12}\left( \boldsymbol{\theta }^{\prime }\right) \\ 
\boldsymbol{I}_{21}\left( \boldsymbol{\theta }^{\prime }\right) & 
\boldsymbol{I}_{22}\left( \boldsymbol{\theta }^{\prime }\right)%
\end{array}%
\right). \label{eqnI12}
\end{align}%

As discussed in
Section \ref{sec3.1}, when $\boldsymbol{I}_{12}\left( \boldsymbol{\theta }%
^{\prime }\right) \neq \boldsymbol{0}$, the influence function $U_{i}\left( \boldsymbol{\theta }^{\prime
}\right) $ alone is not efficient. In this case, valid inference for $\mu$ requires joint estimation of $\boldsymbol{\theta }^{\prime }=(\mu ,\boldsymbol{\psi }^{\prime \top })^{\top}$ to account for variational dependence between $\mu$ and $\boldsymbol{\psi }^{\prime }$.

Let $\widehat{\boldsymbol{\theta }}^{\prime }=( \widehat{\mu },\widehat{%
\boldsymbol{\psi }}^{\prime \top }) ^{\top }$ be the estimator of $%
\boldsymbol{\theta }^{\prime}$ from solving the joint estimating equations $\sum_{i=1}^{n}\boldsymbol{W}_{i}\left( \boldsymbol{\theta }^{\prime }\right)
 =\mathbf{0}$.
Analogous to Section \ref{sec3.3},  $\widehat{\boldsymbol{\theta }}^{\prime }$ is
asymptotically normal with influence function $\boldsymbol{\varphi }_{i}\left( 
\boldsymbol{\theta }^{\prime }\right) =\boldsymbol{I}^{-1}\left( 
\boldsymbol{\theta }^{\prime }\right) \boldsymbol{W}_{i}\left( 
\boldsymbol{\theta }^{\prime }\right) $. Define $\boldsymbol{\theta }_{0}^{\prime }=(\mu_{0} ,\boldsymbol{\psi }_0^{\prime \top })^{\top}$, where $\boldsymbol{\psi }_0^{\prime}$ denotes the probability limit of the estimator $\widehat{%
\boldsymbol{\psi }}^{\prime}$ under the misspecified PS model, the asymptotic variance of $\widehat{\boldsymbol{\theta }}^{\prime }$
thus has the sandwich form:\ 
\begin{equation*}
\Sigma _{\boldsymbol{\theta }}=Var\{\boldsymbol{\varphi }_{i}\left( 
\boldsymbol{\theta }_{0}^{\prime }\right) \} =\boldsymbol{I}^{-1}\left( 
\boldsymbol{\theta }_{0}^{\prime }\right) Var\{ \boldsymbol{W}_{i}\left( 
\boldsymbol{\theta }_{0}^{\prime }\right) \} \boldsymbol{I}^{-1}\left( 
\boldsymbol{\theta }_{0}^{\prime }\right) =\left( 
\begin{array}{cc}
\Sigma _{\mu } & \Sigma _{\mu \boldsymbol{\psi }^{\prime}}
\\ 
\Sigma _{\boldsymbol{\psi }^{\prime}\mu } & \Sigma _{\boldsymbol{\psi }^{\prime}}%
\end{array}%
\right).
\end{equation*}%
Since jointly, $%
\boldsymbol{W}_{i}\left( \boldsymbol{\theta }^{\prime }\right) $ is not
a score function, $\Sigma _{\boldsymbol{\theta }}\neq \boldsymbol{I}^{-1}\left( \boldsymbol{\theta }_{0}^{\prime }\right)$. In particular, the asymptotic variance of $\widehat{\mu }$ is given by $\Sigma _{\mu }$, and because $%
U_{i}\left( \boldsymbol{\theta }^{\prime }\right) $ is an influence function rather than a score, we have $\Sigma _{\mu }\neq I%
_{11}\left( \boldsymbol{\theta }_{0}^{\prime }\right)^{-1} $. \ 

By contrast, when the nuisance parameter is correctly specified so that $\boldsymbol{\psi }^{\prime }=\boldsymbol{\psi }$, the function $U_{i}\left(\boldsymbol{\theta }\right)$ is the efficient influence function (EIF). In this case, with a consistent estimator $\widehat{\boldsymbol{\psi }}$, both $\mu$ and its asymptotic variance can be validly estimated using $U_{i}\left(\boldsymbol{\theta }\right)$ alone, as discussed in Section \ref{sec3.3}.

\subsection{Bootstrap Joint Inference \label{sec4.2}}

As indicated by the simulations in \citealt{shook2025double}, bootstrap
also provides valid inference under misspecified PS, or $\boldsymbol{\psi }%
^{\prime }$. This is not unexpected,\ as we now show that bootstrap
essentially replicates the joint inference procedure of Section \ref{sec3.1}
for each bootstrapped sample. \ 

Consider a bootstrap of size $M$. For the $m$th replicate $Z_{i}^{\left(
m\right) }$ $\left( 1\leq m\leq M\right) $, to estimate $\boldsymbol{\theta }^{\prime }=\left( \mu , \boldsymbol{\psi }^{\prime  \top
}\right){^{\top }}$, one
first solves the $\boldsymbol{V}_{i}$-based estimating equations: 
\begin{equation*}
\sum_{i=1}^{n}\boldsymbol{V}_{i}\left( \boldsymbol{\psi }^{\prime }\right)
=\sum_{i=1}^{n}\boldsymbol{V}\left( Z_{i}^{\left( m\right) },\boldsymbol{\psi }^{\prime
}\right) =\mathbf{0}  
\end{equation*}%
to obtain a nuisance estimator $\widehat{\boldsymbol{\psi }}_{\left( m\right)
}^{\prime }$. Substituting $\widehat{\boldsymbol{\psi }}_{\left( m\right)
}^{\prime }$ into the $U_{i}$-based estimating equation,
\begin{equation*}
\sum_{i=1}^{n} U_{i}( \mu ,\widehat{\boldsymbol{\psi }}_{\left( m\right)
}^{\prime }) =\sum_{i=1}^{n}U( Z_{i},\mu ,\widehat{\boldsymbol{%
\psi }}_{\left( m\right) }^{\prime }) =0
\end{equation*}%
yields $\widehat{\mu}_{\left( m\right) }(\widehat{\boldsymbol{\psi }}_{\left( m\right) }^{\prime })$ for the target parameter, which is a function of the estimated nuisance parameter $\widehat{\boldsymbol{\psi }}_{\left( m\right) }^{\prime }$. Consequently, each bootstrap replicate gives an estimator $\widehat{%
\boldsymbol{\theta }}^{\prime}_{\left( m\right) }=\left( \widehat{\mu}_{\left(
m\right) }, \widehat{\boldsymbol{\psi }}_{\left( m\right) }^{\prime \top }\right) ^{^{\top }}$, which essentially solves the joint estimating equations defined by $\boldsymbol{W}_{i}$ in (\ref{eqn200}). Therefore, the bootstrap procedure inherently accounts for the empirical covariation between $U_{i}$ and $\boldsymbol{V}_{i}$ when estimating the sampling distribution of $\widehat{\mu}$. 

For the bootstrap variance estimation of $\boldsymbol{\theta }^{\prime }$, the estimator is given by the sample variance of $\widehat{\boldsymbol{\theta }^{\prime}}_{\left( m\right) }$:\ 
\begin{equation*}
\widehat{\Sigma }_{\boldsymbol{\theta} ^{\prime }}^{B}=\frac{1}{M}\sum_{m=1}^{M}\left( 
\widehat{\boldsymbol{\theta }^{\prime}}_{\left( m\right) }-\overline{\widehat{%
\boldsymbol{\theta }^{\prime}}}\right) \left( \widehat{\boldsymbol{\theta }^{\prime}}_{\left(
m\right) }-\overline{\widehat{\boldsymbol{\theta }}^{\prime}}\right) ^{\top },
\end{equation*}%
where $\overline{\widehat{\boldsymbol{\theta }^{\prime}}}=1/M\sum_{m=1}^{M}\widehat{%
\boldsymbol{\theta }^{\prime}}_{\left( m\right) }$. Partitioning $\widehat{\Sigma }_{\boldsymbol{\theta}
^{\prime }}^{B}$ according to $\mu $ and $\boldsymbol{\psi }%
^{\prime }$ yields: 
\begin{equation*}
\widehat{\Sigma }_{\boldsymbol{\theta} ^{\prime }}^{B}=\left( 
\begin{array}{cc}
\widehat{\Sigma }_{\mu }^{B} & \widehat{\Sigma }_{\mu \boldsymbol{\psi} ^{\prime }}^{B}
\\ 
\widehat{\Sigma }_{\boldsymbol{\psi} ^{\prime } \mu}^{B} & \widehat{\Sigma }_{\boldsymbol{\psi} ^{\prime }}^{B}%
\end{array}%
\right).
\end{equation*}%
The marginal component $\widehat{\Sigma }_{\mu }^{B}$ consistently estimates the variance of $\widehat{\mu }$, since $\widehat{\Sigma }_{\boldsymbol{\theta}
^{\prime }}^{B}$ is consistent for $%
\Sigma _{\boldsymbol{\theta} ^{\prime }}$ 
by standard properties of bootstrap inference \citep{bickel1981some}. \ 

Thus, each bootstrap sample preserves the empirical covariation between $U_{i}$ and $\boldsymbol{V}_{i}$, which ensures that $\widehat{\Sigma }_{\boldsymbol{\theta }^{\prime }}^{B}$ captures the variational dependence between $\mu$ and $\boldsymbol{\psi }^{\prime}$. Although the bootstrap procedure appears to follow the plug-in approach when calculating the variance estimate $\widehat{\Sigma }_{\mu }^{B}$, it in fact performs implicit joint inference, thereby correctly accounting for the covariation between $U_{i}$ and $\boldsymbol{V}_{i}$.

\section{Sample Splitting and Cross Fitting (SSCF)\label{sec4.4}}

We have discussed several strategies for addressing variational dependence between the target and misspecified nuisance parameters by accounting for the covariation between their estimating functions, $\boldsymbol{U}_i$ and $\boldsymbol{V}_i$. These strategies include joint inference approaches (Sections \ref{sec3.1}, \ref{sec4.1}) and the bootstrap procedure that implicitly performs joint inference (Section \ref{sec4.2}), both empirically validated by \citealt{shook2025double}. We also introduced a method that removes such dependence by constructing the efficient score for $\boldsymbol{\mu}$ (Section \ref{sec3.2}).

We now consider another alternative approach that empirically enforces variational independence for a general setting with a $p$-dimensional vector $\boldsymbol{\mu}$ as the target parameter. The key idea is to split the sample and construct a mixture of estimating functions that are uncorrelated by design, thereby yielding a plug-in variance estimator that remains consistent under nuisance misspecification.

\subsection{Variational Independence via Sample Split}
Let the joint estimating function be
\[
\boldsymbol{W}%
_{i}\left( \boldsymbol{\theta }^{\prime }\right) =\left( \boldsymbol{U}_{i}^{\top } ,\boldsymbol{V}_{i}^{\top } \right) ^{\top }.
\]
We partition the $n$ sample  estimating functions into two halves and define $\left( 1\leq i\leq n\right)$:
\begin{equation*}
\boldsymbol{W}_{ri}\left( \boldsymbol{\theta }^{\prime }\right) = \left \{ 
\begin{array}{ll}
\boldsymbol{W}_{ui}\left( \boldsymbol{\theta }^{\prime }\right) = \left( \boldsymbol{U}_{i}\left( \boldsymbol{\mu},\boldsymbol{\psi }^{\prime }\right)^{\top } ,\boldsymbol{0}^{\top } \right) ^{\top }, & \text{for }r=u, \\ 
\boldsymbol{W}_{vi}\left( \boldsymbol{\theta }^{\prime }\right) =\left(\boldsymbol{0}^{\top }, \boldsymbol{V}_{i}\left( \boldsymbol{\psi }^{\prime }\right)^{\top } \right)^{\top }, & \text{for }r=v,%
\end{array} 
\right.  
\end{equation*}%
where each estimating function is assigned to $r=u$ or $r=v$ with equal probability $1/2$.
Let $F_u(\boldsymbol{\theta}')$ and $F_v(\boldsymbol{\theta}')$ denote the distributions of $\boldsymbol{W}_{ui}$ and $\boldsymbol{W}_{vi}$, respectively. Then $\boldsymbol{W}_{ri}(\boldsymbol{\theta}')$ can be viewed as an i.i.d. sample from the 50:50 mixture: \[F_{uv}\left( \boldsymbol{\theta }^{\prime }\right) =\frac{1}{2%
}F_{u}\left( \boldsymbol{\theta }^{\prime }\right) +\frac{1}{2}F_{v}\left( 
\boldsymbol{\theta }^{\prime }\right), \]
whose mean and variance are:\ 
\begin{align}
E\{ \boldsymbol{W}_{ri}\left( \boldsymbol{\theta }^{\prime }\right) \} & =%
\frac{1}{2}\{ E\left( \boldsymbol{W}_{ui}\right) +E\left( \boldsymbol{W}_{vi}\right) \} =\mathbf{0},
\notag \\
Var\{ \boldsymbol{W}_{ri}\left( \boldsymbol{\theta }^{\prime }\right) \} & =%
\frac{1}{2}Var\left( \boldsymbol{W}_{ui}\right) +\frac{1}{2}Var\left( \boldsymbol{W}_{vi}\right) =%
\frac{1}{2}\left( 
\begin{array}{cc}
Var\left( \boldsymbol{U}_{i}\right) & \boldsymbol{0} \\ 
\boldsymbol{0} & Var\left( \boldsymbol{V}_{i}\right)%
\end{array}%
\right).  \label{eqn502}
\end{align}%
By construction, $\boldsymbol{W}_{ri}\left( \boldsymbol{\theta }^{\prime }\right) $ is a mean-zero estimating
function for $\boldsymbol{\theta }^{\prime }$, and the off-diagonal block terms of \eqref{eqn502} vanish, which ensures that $\boldsymbol{\mu}$ and $\boldsymbol{\psi}'$ are variationally independent under $\boldsymbol{W}_{ri}(\boldsymbol{\theta}')$. We then estimate $\boldsymbol{\theta}'$ by solving 
\begin{equation}
\sum_{i=1}^{n}\boldsymbol{W}_{ri}(\boldsymbol{\theta}')=\mathbf{0}. \label{eqn510}
\end{equation}
Further, if $\boldsymbol{U}_{i}\left( \boldsymbol{\mu},\boldsymbol{\psi }^{\prime }\right)$ is an influence function and $\boldsymbol{V}_{i}\left( \boldsymbol{\psi }^{\prime }\right) $ is a mean-zero estimating function in general, it follows from Section %
\ref{sec3.3} that the \textquotedblleft Fisher information" can be defined as: 
\begin{equation}
\boldsymbol{I}\left( \boldsymbol{\theta }^{\prime }\right) =E\{-\frac{\partial }{%
\partial \boldsymbol{\theta } }\boldsymbol{W}_{ri}\left( \boldsymbol{\theta }^{\prime }\right) %
\} =\frac{1}{2}\left( 
\begin{array}{cc}
\mathbf{I}_{p} & \boldsymbol{0} \\ 
\boldsymbol{0} & E\left\{\boldsymbol{\ V}_{i}\left( \boldsymbol{\theta }%
\right) \boldsymbol{S}_{\boldsymbol{\psi },i}^{\top }\left( \boldsymbol{%
\theta }\right) \right \}%
\end{array}%
\right).  \label{eqn504}
\end{equation}
Hence, $\widehat{\boldsymbol{\theta}}^{\prime
}$ is asymptotically linear \citep{hampel1974influence,tsiatis2006semiparametric}:
\begin{align}
\sqrt{n}\left( \widehat{\boldsymbol{\theta }}^{\prime }-\boldsymbol{\theta }_0%
^{\prime }\right) & =\frac{\sqrt{n}}{n}\sum_{i=1}^{n}\boldsymbol{I}^{-1}\left( 
\boldsymbol{\theta }_0^{\prime }\right) \boldsymbol{W}_{ri}+\mathbf{o}_{p}\left(
1\right)\rightarrow _{d}N\left( 0,\Sigma _{\boldsymbol{\theta} ^{\prime
}}\right),  \label{eqn520} 
\end{align}
whose variance can be calculated from (\ref{eqn502}) and (\ref{eqn504}): 
\begin{equation*}
\Sigma _{\boldsymbol{\theta} ^{\prime }}=\boldsymbol{I}^{-1}\left( \boldsymbol{\theta }_0^{\prime
}\right) Var\left( \boldsymbol{W}_{ri}\right) \boldsymbol{I}^{-1}\left( \boldsymbol{\theta }_0%
^{\prime }\right) =2\left( 
\begin{array}{cc}
Var\left \{ \boldsymbol{U}_{i}\left( \boldsymbol{\theta }_0^{\prime }\right) \right \}
& \mathbf{0} \\ 
\mathbf{0} & E^{-1}\left\{\boldsymbol{\ V}_{i}\left( \boldsymbol{\theta }_0^{\prime
}%
\right) \boldsymbol{S}_{\boldsymbol{\psi },i}^{\top }\left( \boldsymbol{%
\theta }_0^{\prime
}\right) \right \}%
\end{array}%
\right).  
\end{equation*}%
Consequently, the target parameter readily satisfies:
\begin{equation}
\sqrt{n}\left( \widehat{\boldsymbol{\mu} }-\boldsymbol{\mu} _{0}\right) \rightarrow _{p}N\left( \boldsymbol{0},2Var%
\{\boldsymbol{ U}_{i}\left( \boldsymbol{\theta }_0^{\prime }\right) \} \right).
\label{eqn550}
\end{equation}

This estimator in \eqref{eqn550} essentially uses only half of the $\boldsymbol{U}_i$-based and half of the $\boldsymbol{V}_i$-based estimating functions (or close to half if $n$ is not an integer), which is consistent but inefficient, with an asymptotic variance that is twice that of the full-sample estimator. In practice, we reverse their roles and re-estimate on the complementary split, motivating the cross-fitting step. 

\subsection{Implementation}
To implement the above procedure, we first create two i.i.d.
samples from the mixture $F_{uv}\left( \boldsymbol{\theta }^{\prime
}\right) $ by randomly partitioning the original sample $\left \{ \boldsymbol{W}_{i}\left( 
\boldsymbol{\theta }^{\prime }\right) ,1\leq i\leq n\right \} $ into two complementary subsets of sizes $n_1$ and $n_2$, respectively. Specifically, we assign half of the $\boldsymbol{U}_i$- and $\boldsymbol{V}_i$-based estimating functions to form Sample 1, and swap their roles to form Sample 2.

For instance, $\left \{ \boldsymbol{W}_{ui},1\leq i\leq n_{1}\right \} $ and $\left \{
\boldsymbol{W}_{vj},n_{1}+1\leq j \leq n\right \} $ belong to the Sample 1 and $\left \{ \boldsymbol{W}_{ui},n_{1}+1\leq i\leq n\right \} $ and $\left \{
\boldsymbol{W}_{vj},1\leq j \leq n_{1}\right \} $ form the Sample 2, yielding:  
\begin{align}
\text{Sample 1}& :\boldsymbol{W}_{u1},\boldsymbol{W}_{u2},\ldots ,\boldsymbol{W}_{un_{1}},\boldsymbol{W}_{v\left( n_{1}+1\right)
},\boldsymbol{W}_{v\left( n_{1}+2\right) },\ldots ,\boldsymbol{W}_{vn},  \label{eqn560} \\
\text{Sample 2}& :\boldsymbol{W}_{v1},\boldsymbol{W}_{v2},\ldots ,\boldsymbol{W}_{vn_{1}},\boldsymbol{W}_{u\left( n_{1}+1\right)
},\boldsymbol{W}_{u\left( n_{1}+2\right) },\ldots ,\boldsymbol{W}_{un}.  \notag
\end{align}%
Within each sample, the estimating functions are i.i.d.; between the two samples, they are correlated due to shared dependence on the original data. Because $\boldsymbol{W}_{ri}$ ($1\leq i\leq n$) is a 50:50 mixture of $\boldsymbol{W}_{ui}$ and $\boldsymbol{W}_{vi}$, it follows that $\lim_{n\rightarrow
\infty }\frac{n_{1}}{n}=\lim_{n\rightarrow \infty }\frac{n_{2}}{n}=\frac{1}{2%
}$.

\begin{figure}[!htbp]
\centerline{\includegraphics[width=6in]{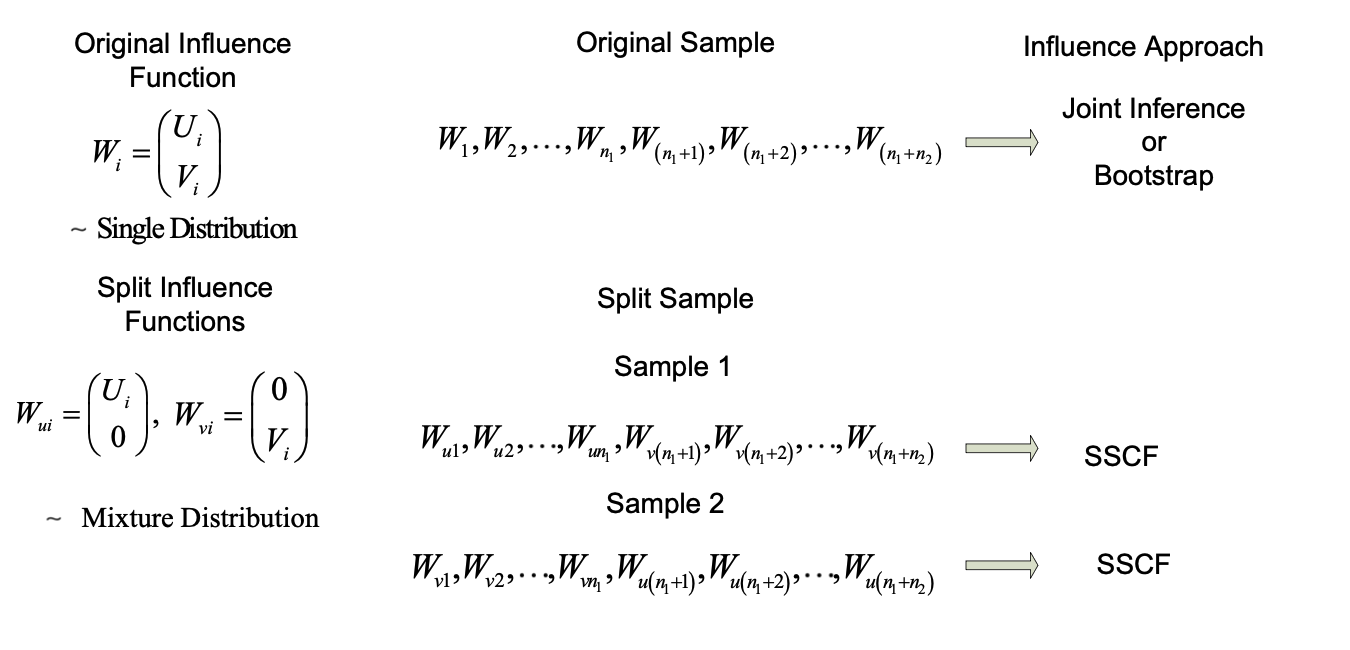}} \vspace*{-3pt}
\caption{Sample Splitting and Cross Fitting of Influence Functions to Break
Variational Dependence}
\label{fig:figure1}
\end{figure}

Figure $\ref{fig:figure1}$ shows how two complementary mixture samples are constructed from the original sample and how this procedure differs from joint inference and bootstrap. To estimate the target parameter $\boldsymbol{\mu}$, we proceed as follows.

First, using Sample 1 in \eqref{eqn560}, we solve (\ref{eqn510}) for $%
\boldsymbol{\theta }^{\prime }$. Essentially, this is the same as
first solving the $\boldsymbol{V}_{j}$-based estimating equations for $\boldsymbol{\psi }%
^{\prime }$ to obtain an estimator $\widehat{\boldsymbol{\psi }}_{1}^{\prime }$: 
\begin{equation*}
\sum_{j=n_{1}+1}^{n}\boldsymbol{V}_{j}\left( \boldsymbol{\psi }^{\prime }\right) =\mathbf{%
0},  
\end{equation*}%
and then plug in $\widehat{\boldsymbol{\psi }}_{1}^{\prime
}$ and solve the $\boldsymbol{U}_{i}$-based equations for $\boldsymbol{\mu} $ to obtain $\widehat{\boldsymbol{\mu} }_{1}$: 
\begin{equation*}
\sum_{i=1}^{n_{1}}\boldsymbol{U}_{i}( \boldsymbol{\mu} ,\widehat{\boldsymbol{\psi }}_{1}^{\prime
}) =\mathbf{0}.  
\end{equation*}%
The asymptotic variance of $\widehat{\boldsymbol{\mu} }_{1}$ is estimated by the plug-in estimator:%
\begin{equation*}
\widehat{\Sigma }_{\boldsymbol{\mu} ,1}=\frac{1}{n_{1}}\sum_{i=1}^{n_{1}}\boldsymbol{U}_{i}( 
\widehat{\boldsymbol{\mu} }_{1},\widehat{\boldsymbol{\psi }}_{1}^{\prime })\boldsymbol{U}_{i}^{\top}( 
\widehat{\boldsymbol{\mu} }_{1},\widehat{\boldsymbol{\psi }}_{1}^{\prime }).
\end{equation*}

Next, applying the same steps to Sample 2 yields $\widehat{\boldsymbol{\mu} }_{2}$ and $\widehat{\Sigma }_{\boldsymbol{\mu},2}$. \ We then average the two sets of estimators to improve efficiency.

\begin{theorem} 
\label{thm1}
    Let $\widehat{\boldsymbol{\mu}}_{k}$ and $\widehat{\Sigma}_{\boldsymbol{\mu},k}$ denote the two sets
of cross-fitted estimators from the two samples in (\ref{eqn560}) $\left( k=1,2\right) $.
\ Define the combined point and variance estimator as: 
\begin{equation*}
\widehat{\boldsymbol{\mu}}^{sscf}=\frac{1}{2}\left( \widehat{\boldsymbol{\mu}}_{1}+\widehat{\boldsymbol{\mu}}_{2}\right) ,%
\mathbf{\quad}\widehat{\Sigma}_{\boldsymbol{\mu}}^{sscf}=\frac{1}{2}\left( \widehat{\Sigma}%
_{\boldsymbol{\mu},1}+\widehat{\Sigma}_{\boldsymbol{\boldsymbol{\mu}},2}\right)   
\end{equation*}
Under mild regularity conditions, denote by $\rightarrow_{p}$ convergence in probability, we have
\begin{enumerate}
    \item $\widehat{\boldsymbol{\mu}}^{sscf} \rightarrow_{p} \boldsymbol{\mu}_0$, the ground truth;
    \item $\widehat{\Sigma}^{sscf}_{\boldsymbol{\mu}}\rightarrow_{p} \Sigma_{\boldsymbol{\mu}}$ for  the sample-splitting and cross-fitting (SSCF) estimator $\widehat{\boldsymbol{\mu}}^{sscf}$.
\end{enumerate}
\end{theorem}

Thus, by estimating the nuisance $\boldsymbol{\psi }^{\prime }$ on one partition and the target $\boldsymbol{\mu} $ on the complementary partition,
we can, again, obtain a plug-in variance estimator that remains consistent even when the nuisance model is misspecified. This procedure corresponds to the sample-splitting and cross-fitting (SSCF) framework discussed in the literature %
(\citealt{robins2008higher}, \citealt{chernozhukov2018double}). 
Although consistency of $\widehat{\Sigma}_{\boldsymbol{\mu},k}$ is expected, $\widehat{\Sigma}^{sscf}_{\boldsymbol{\mu}}=1/2\left( \widehat{\Sigma}%
_{\boldsymbol{\mu},1}+\widehat{\Sigma}_{\boldsymbol{\boldsymbol{\mu}},2}\right)$ should not be taken for granted, as each $\widehat{\boldsymbol{\mu}}_k$ depends on a nuisance estimated from the complementary subsample. A formal proof of Theorem \ref{thm1} is provided in the Supplementary Material Web Appendix B.

\section{Applications\label{sec5}}

\subsection{Simulation Study}

We conducted a Monte Carlo (MC) simulation study to illustrate (1)\ the covariation
between $U_{i}(\mu,\boldsymbol{\psi },\boldsymbol{\xi }%
)$, the influence function (IF) of the doubly robust estimator (DRE), and $\boldsymbol{V}_{i}\left( \boldsymbol{\psi }^{\prime }\right)$, the score function for $\boldsymbol{\psi }^{\prime }$ under a misspecified PS
model; and (2) the performance of the sample-splitting and cross-fitting (SSCF) approach in estimating the asymptotic variance. Throughout, the outcome regression model was assumed to be correctly specified. We adapted the simulation setting in \citealt{shook2025double},
with sample size $n=800$ and $M=5000$ MC replications, but modified to our objectives. \ 

\subsubsection{Data Generating Process (DGP)\label{sec5.1}}

We generated three covariates $\mathbf{z}%
_{i}=\{ z_{i1},z_{i2},z_{i3}\} ^{\top }$: 
$z_{i1}\sim N\left( 5,4\right) ,\text{ } z_{i2}\sim \text{Bernoulli}(0.25),\text{ }
z_{i3}\sim \text{Bernoulli}(0.75).$ The propoensity score (PS) for the binary treatment $x$ was simulated from: 
$$
\begin{aligned}
x_{i}& \mid \mathbf{z}_{i}\sim \text{Bernoulli}(\eta _{i}),\quad \eta
_{i}=E(x_{i}\mid \mathbf{z}_{i}),  \label{eqn570_20} \\
\eta _{i}\left( \boldsymbol{\psi }_{0}\right) & =\eta \left( \mathbf{z}_{i};%
\boldsymbol{\psi }_{0}\right) = \text{expit}\left(
0.5+0.5z_{i2}-0.2z_{i1}z_{i2}\right).  \notag
\end{aligned}
$$
For each treatment $x_{i}=k$ $\left( =0,1\right) $, the potential outcome $%
y_{i}^{k}$ follows:%
\begin{align*}
y_{i}^{k}& \mid \mathbf{z}_{i}\sim N(\mu _{i}^{k}\left( \mathbf{z}%
_{i}\right) ,\sigma ^{2}),\quad \mu _{i}^{k}\left( \mathbf{z}_{i}\right)
=E(y_{i}^{k}\mid x_{i}=k,\mathbf{z}_{i}),\quad \sigma =400,\quad \\
\mu _{i}^{k}\left( \mathbf{z}_{i}\right) &
=1000+11.5z_{i1}+100z_{i2}-15z_{i1}z_{i2}+25k-5.5kz_{i1}-30kz_{i2}+5kz_{i1}z_{i2}.
\end{align*}

The observed outcome was defined as: $%
y_{i}=x_{i}y_{i}^{1}+(1-x_{i})y_{i}^{0} $, and the true average causal effect (ACE)
is: 
\begin{equation*}
\mu _{0}=E\left( y_{i}^{1}-y_{i}^{0}\right) =E\{ \mu _{i}^{1}\left( 
\mathbf{z}_{i}\right) -\mu _{i}^{0}\left( \mathbf{z}_{i}\right) \}
\end{equation*}%
Under the above simulation setting, this true ACE$\  \mu _{0}$ was estimated
to be 15.02 by a large MC sample size $M_{true}=5\times 10^{7}$. \ 

As noted earlier, we assumed that the OR model was correctly specified,
which was given by a parametric linear regression with the conditional mean:\ 
\begin{equation*}
Q_{i}\left( x_{i};\mathbf{z}_{i},\boldsymbol{\xi }\right) =\xi _{0}+\xi
_{1}z_{i1}+\xi _{2}z_{i2}+\xi _{3}z_{i1}z_{i2}+\xi _{4}x_{i}+\xi
_{5}x_{i}z_{i1}+\xi _{6}x_{i}z_{i2}+\xi _{7}x_{i}z_{i1}z_{i2}.
\end{equation*}%
Additionally, we assumed a correctly specified PS $\boldsymbol{\psi }$ as
the data generating process of PS: 
\begin{equation*}
\eta _{i}\left( \boldsymbol{\psi }\right) =\eta \left( \mathbf{z}_{i},%
\boldsymbol{\psi }\right) =\text{expit}\left( \psi _{0}+\psi _{1}z_{i2}+\psi
_{2}z_{i1}z_{i2}\right) ,  
\end{equation*}%
we consider a misspecified PS $\boldsymbol{\psi }^{\prime }$ including only the
intercept and $\sin (z_{i1})$ as $\eta _{i}^{\prime }\left( \boldsymbol{\psi }^{\prime }\right) =\eta ^{\prime
}\left( \mathbf{z}_{i},\boldsymbol{\psi }^{\prime }\right) =\text{expit}%
\{\psi _{0}^{\prime }+\psi _{1}^{\prime }\sin (z_{i1})\}$. We provide inference procedures for both
correctly specified and misspecified nuisance in the Supplementary Material Appendix C.

\subsubsection{Covariation between DRE\ and Nuisance Estimating Functions}

We used MC simulations to demonstrate that under
the correctly-specified PS model $\boldsymbol{\psi }$, the cross-information term in (\ref{eqnI12}), $\boldsymbol{I}_{12}\left( \boldsymbol{\theta }%
_{0}\right) =E\{ U_{i}\left( \boldsymbol{\theta }_{0}\right) \boldsymbol{V}%
_{i}^{\top }\left( \boldsymbol{\psi }_{0}\right) \} =\mathbf{0}$, whereas under
a misspecified PS model $\boldsymbol{\psi }^{\prime }$, this term $\boldsymbol{I}_{12}\left( 
\boldsymbol{\theta }_{0}^{\prime }\right) =E\{ U_{i}( 
\boldsymbol{\theta }_{0}^{\prime }) \boldsymbol{V}_{i}^{\top 
}\left( \boldsymbol{\psi }_{0}^{\prime }\right) \} \neq 
\mathbf{0}$. We
estimated the correlations between $U_{i}\left( \boldsymbol{\theta }%
_{0}\right) $ and each component of $\boldsymbol{V}_{i}\left( \boldsymbol{\psi }%
_{0}\right) $ using the sample Pearson correlation between $U_{i}( \widehat{%
\boldsymbol{\theta }}) $ and each component of $\boldsymbol{V}_{i}( \widehat{%
\boldsymbol{\psi }}) $ under the correctly specified PS $\boldsymbol{%
\psi }$. Likewise, under the misspecified PS $%
\boldsymbol{\psi }^{\prime }$, we estimated their correlations using the sample
Pearson correlation between $U_{i}( \widehat{\boldsymbol{%
\theta }}^{\prime }) $ and each component of $\boldsymbol{V}_{i}( \widehat{%
\boldsymbol{\psi }}^{\prime }) $ . \ 

Shown in Table \ref{tab:pearson} are estimated correlations between $U_{i}$ and $%
\boldsymbol{V}_{i}$ for the two scenarios based on MC replication. All the correlations
were close to zero under the correctly specified PS. In contrast, with the
misspecified PS $\boldsymbol{\psi }^{\prime }$, the correlations between $%
U_{i}^{\prime }$ and $\boldsymbol{V}_{i}^{\prime }$ deviated from zero by nearly 10 times
larger than their counterparts under the correctly-specified PS. \textbf{\ }


\subsubsection{Comparison of Asymptotic Variance Estimators}

The simulation study results in \citealt{shook2025double} demonstrated bias
in variance estimation of $\widehat{\mu }$ based on the variance
of $U_{i}\left( \boldsymbol{\theta }^{\prime }\right) $ using the
plug-in rule, when PS $\boldsymbol{\psi }^{\prime }$ is misspecified and OR $\boldsymbol{\xi }$ is correctly specified. Building on these findings, we further compared the $U_{i}^{\prime }$%
-based variance estimators obtained via the plug-in rule, with and without the sample-splitting and cross-fitting (SSCF) discussed in Section \ref{sec4.4}. To compare the accuracy, we calculated the standard error ratio
(SER), defined as the ratio of the square root of the variance estimate from
each method to that from the Monte Carlo. \ A value of SER\ closer to 1
indicates less bias, while a value of SER larger or smaller than 1 suggests
over- or under-estimation of the asymptotic variance. \ 

Without SSCF, the plug-in variance estimator underestimated the true
variability of $\widehat{\mu }$, with a standard error of 28.7 compared to
the Monte Carlo estimate of 29.5 (SER = 0.97), which is consistent with
downward bias observed in the simulation results by \citealt{shook2025double}. In contrast, SSCF corrected this bias; the plug-in and Monte
Carlo estimates were nearly identical, 29.2 vs. 29.3, respectively, yielding
an SER of 1.00. These results indicate that SSCF improves the accuracy of
variance estimation. \ 

\subsection{Real Data}

For illustration with a real study data, we used a labor training program (%
\citealt{lalonde1986evaluating}) from the National Supported Work
Demonstration, which has been previously used to assess the causal effect of
training participation on post-program earnings. Among the 614
subjects, 185 are in the treatment and 429 are in the control group. Baseline
covariates include age, education, race, marital status, an indicator for
lacking a high school diploma, and pre-intervention earnings in 1974 and
1975. The treatment variable is binary (1 = treated, 0 = control), and the
outcome is the continuous, post-training real earnings in 1978. \ 

In our analysis, we included all aforementioned covariates when estimating the OR model to approximate the true data generating process. To evaluate inference under a misspecified PS model, we intentionally restricted the PS model to include only age, education, race, and marital status, omitting other important covariates. While this specification increases the likelihood of PS misspecification, the OR model may also be misspecified because the true data generating process is unknown. Nonetheless, all variance estimation methods examined in this paper, except for the IF-based variance estimator, will retain double robustness. That is, under mild regularity conditions (e.g., asymptotic linearity), they continue to yield valid variance estimates for the DRE when one nuisance model is misspecified. Moreover, even when both the OR and PS models are incorrect, these methods remain consistent for variance estimation, although the DRE itself may be biased in point estimation (\citealt{shook2025double}).

Shown in \textbf{Table~\ref{tab:psmodel}} is a summary of the PS specifications
and correlations for assessing covariation between the influence function of
the DRE\ and the score function of the specified PS model. These correlation values, even though not large in scale, can lead to large bias in variance estimation if we use IF-based variance estimation without SSCF. 

Shown in \textbf{Table~\ref{tab:se_comparison}} are standard error estimates: the DRE IF-based plug-in
standard errors with and without SSCF, along with the jointly estimated
standard errors. As noted earlier, the jointly estimated standard error is used as the
benchmark because it remains valid regardless of nuisance parameter
specification. The results show that the DRE is correlated with the score
function of the specified PS. Without SSCF, the DRE IF-based plug-in
estimator substantially underestimated the standard error (775.8). \
Incorporating SSCF yielded an estimate (821.6) much closer to the jointly
estimated benchmark (816.0). \ 


\section{Discussion\label{sec6}}

In this paper, we addressed an important but under-discussed question: Why is the double-robust estimator
(DRE) for causal inference not doubly robust for variance estimation?
Building on the simulation results of \citealt{shook2025double}, we showed
that bias in the IF-based variance estimator arises because the influence
function (IF) of DRE\ is only efficient under correctly specified nuisance
parameter models. This failure reflects the variational dependence, or
covariation, between the target misspecified nuisance parameters and their corresponding estimating functions. \ 

Our findings clarify an important distinction from classical semiparametric
theory. Efficient influence functions in generalized estimating equations
(GEEs) remain valid for a broad class of nuisance specifications because
they are constructed to be orthogonal to an infinite-dimensional nuisance
tangent space. By contrast, the IF of DRE\ is efficient only under correct
parametric nuisance models. When either the propensity score (PS) or outcome
regression (OR) is misspecified, efficiency fails and variance estimators
become inconsistent. These insights extend earlier discussions (\citealt{munoz2012population}) while also explaining the simulation results from %
\citealt{shook2025double}.

We further discussed several strategies for achieving valid inference under nuisance misspecification. The first is by orthogonalizing  the estimating function with respect to nuisance scores to restore efficiency. The second uses sample-splitting and cross-fitting (SSCF) to construct variationally independent estimating functions for the target and nuisance parameters. We further provide a mixture-distribution justification for this SSCF procedure popularized in the machine learning literature, which, to our knowledge, offers the first explicit mixture-based rationale for its validity under nuisance misspecification. Finally, both joint inference and the bootstrap naturally account for the induced variational dependence and thus yield valid inference.

A comparison of the four approaches illustrates clear trade-offs. The
bootstrap is simple but computationally intensive. Joint inference requires
estimating the joint asymptotic variance of the estimator for both the
target and nuisance parameters. Constructing the efficient estimating function
requires additional programming, but provides a transparent view of the magnitude of covariation. The SSCF, by far,  is the easiest to implement, and paves the way for future integration with machine learning methods. \ 

In summary, we elucidated why the DRE variance estimator fails to maintain
double robustness and how different alternatives, including the ones
introduced in this paper, work to provide consistent variance estimation. These
clarifications will strengthen the theoretical foundation of semiparametric causal
inference and offer practical directions for robust applications where
nuisance models are inevitably imperfect in this real world. \ 

\section*{Acknowledgements}

We thank the Co-Editor and anonymous reviewers for their constructive comments and helpful suggestions that improved this manuscript. We also appreciate the support and feedback from our colleagues. The content is solely the responsibility of the authors and does not necessarily reflect the official views of the supporting institutions.\vspace*{-10pt}


\section*{Supplementary Materials}

Web appendices in Section 3, 5 and 6 are available with this paper at the Biometrics website on
Wiley Online Library.\vspace*{-10pt}

\bibliographystyle{biom}
\bibliography{biomsample}

\newpage
\section*{Supplementary Material }
\subsection*{Appendix A. Efficient Estimating Function and Sandwich Variance Estimator}
\subsubsection*{1. Example of Efficient Estimating Functions under Misspecified Nuisance}
As noted in Section 3, in this web appendix, we illustrate by examples that efficient estimating
functions for target parameters for semiparametric, or restricted moment,
models remain efficient with misspecified nuisance (distribution) parameters.

Consider a negative binomial (NB) model for a count response, $y_{i}\sim 
\mathrm{NB}\left( \mu ,\alpha \right) $, with mean $\mu $ and dispersion
parameter $\alpha $. Now suppose that one misspecifies $y_{i}$ to follow $%
\mathrm{Poisson}\left( \mu \right) $. The parameter of interest $\mu $
remains the same, but the nuisance parameters differ between the two models,
with NB having an additional nuisance parameter $\psi =\alpha $. \ 

The Fisher information for $\boldsymbol{\theta }=\left( \mu ,\alpha \right)
^{^{\top }}$ derived from the NB score function $S_{\boldsymbol{\theta} i}(\boldsymbol{%
\theta })=\left( S_{\mu i}(\boldsymbol{\theta }),S_{\alpha i}(\boldsymbol{%
\theta })\right) ^{\top }$ is: 
\begin{equation*}
\boldsymbol{I}\left( \boldsymbol{\theta }\right) =\left( 
\begin{array}{cc}
I_{11}\left( \boldsymbol{\theta }\right) & I_{12}\left( \boldsymbol{\theta }%
\right) \\ 
I_{21}\left( \boldsymbol{\theta }\right) & I_{22}\left( \boldsymbol{\theta }%
\right)%
\end{array}%
\right) =\left( 
\begin{array}{cc}
E\{-\frac{\partial }{\partial \mu }S_{\mu i}(\boldsymbol{\theta })\} & E\{-%
\frac{\partial }{\partial \alpha }S_{\mu i}(\boldsymbol{\theta })\} \\ 
E\{-\frac{\partial }{\partial \alpha }S_{\mu i}(\boldsymbol{\theta })\} & 
E\{-\frac{\partial }{\partial \alpha }S_{\alpha i}(\boldsymbol{\theta })\}%
\end{array}%
\right) .
\end{equation*}%
It is deduced that the off-diagonal term $I_{12}(\boldsymbol{\theta })$, or $%
E\{-\frac{\partial }{\partial \alpha }S_{\mu i}(\boldsymbol{\theta })\}$, is 
$0$, showing no covariation between $S_{\mu i}$ and $S_{\alpha i}$.
Consequently, the score of interest $S_{\mu i}$ is orthogonal to $S_{\alpha
i}$, and hence, efficient, with variance $I_{11}\left( \boldsymbol{\theta }%
\right) =\frac{1}{\mu \left( 1+\alpha \mu \right) }$. 

Let $\widehat{\mu }_{NB}$ denote the estimator obtained by solving the $%
S_{\mu i}$-based score equation: 
\begin{equation}
\sum_{i=1}^{n}S_{\mu i}\left( \boldsymbol{\theta }\right) =\sum_{i=1}^{n}%
\frac{1}{\mu \left( 1+\alpha \mu \right) }\left( y_{i}-\mu \right) =0
\label{eqn90}
\end{equation}%
Then the MLE\ $\widehat{\mu }_{NB}$ is consistent and asymptotically normal,
and its asymptotic variance $\sigma _{NB}^{2}$ is the inverse of Fisher
information $I_{11}\left( \boldsymbol{\theta }\right) $, i.e., 
\begin{equation}
\sigma _{NB}^{2}=I_{11}^{-1}\left( \boldsymbol{\theta }_{0}\right) =\mu
_{0}\left( 1+\alpha _{0}\mu _{0}\right) .\   \label{eqn100}
\end{equation}%
It is also readily shown that the EIF\ for $\widehat{\mu }_{NB}$ is $%
\varphi \left( y_{i},\mu _{0}\right) =\left( y_{i}-\mu _{0}\right) $ and
thus $\sigma _{NB}^{2}=Var\{ \varphi \left( y_{i},\mu _{0}\right)
\} $. \ 

Let $y_{i}^{\prime }\sim \mathrm{Poisson}\left( \mu \right) $. Unlike $y_{i}$%
, $y_{i}^{\prime }$ has $\mathrm{Poisson}\left( \mu \right) $ as its DGP.
The Fisher information is $I^{^{\prime }}\left( \mu \right) =E\{-\frac{%
\partial }{\partial \mu }S_{\mu i}^{\prime }(\mu )\}=E\left( \frac{%
y_{i}^{\prime }}{\mu ^{2}}\right) =\frac{1}{\mu }\nonumber$, where the
Poisson score $S_{\mu i}^{\prime }(\mu )=\frac{\partial }{\partial \mu }%
l_{i}^{\prime }(\mu )=\frac{1}{\mu }\left( y_{i}^{\prime }-\mu \right) $.
Let $\widehat{\mu }_{Poi}$ denote the estimator from solving the score
equation, $\sum_{i=1}^{n}S_{\mu i}^{\prime }(\mu )=0$. Then the Poisson MLE $%
\widehat{\mu }_{Poi}$ has asymptotic variance $\sigma
_{Poi}^{2}=\{I^{^{\prime }}\left( \mu _{0}\right) \}^{-1}=\mu _{0}$. \ 

Since $y_{i}$ is generated from NB, the Poisson score $S_{\mu i}^{\prime
}(y_{i},\mu _{0})$ is misspecified. However, the $S_{\mu i}^{\prime
}$-based score equation: 
\begin{equation}
\sum_{i=1}^{n}S_{\mu i}^{\prime}\left( \boldsymbol{\theta }\right) =\sum_{i=1}^{n}%
\frac{1}{\mu }\left( y_{i}-\mu \right) =0  \label{eqn110}
\end{equation}%
still yields a consistent and asymptotically normal estimator $\widehat{\mu }%
^{\prime }$ for $\mu _{0}$, but the asymptotic variance, $\sigma _{sw}^{2}$,
is no longer equal to the inverse of Fisher information under Poisson, $%
\sigma _{sw}^{2}\neq \mu _{0}$. \ 

To see this, let $B^{\prime }\left( \mu _{0}\right) =E\{-\frac{\partial }{%
\partial \mu }S_{\mu i}^{\prime }(y_{i},\mu _{0})\}$. Since $y_{i}$ does not
follow Poisson, $B^{\prime }\left( \mu _{0}\right) \neq I^{^{\prime }}\left(
\mu _{0}\right) $. From the asymptotic linearity of $\widehat{\mu }%
^{\prime }$ (\citealt{hampel1974influence}, \citealt{tsiatis2006semiparametric}), we have:
\begin{equation*}
\sqrt{n}\left( \widehat{\mu }^{\prime }-\mu _{0}\right) =\frac{\sqrt{n}}{n}%
\sum_{i=1}^{n}\varphi ^{\prime }\left( y_{i},\mu _{0}\right) +\boldsymbol{o}_{p}\left(
1\right) ,
\end{equation*}%
where $\boldsymbol{o}_{p}\left( 1\right) $ denotes the stochastic $\boldsymbol{o}\left( 1\right) $ and 
$\varphi ^{\prime }\left( y_{i},\mu _{0}\right) =\{B^{\prime }\left( \mu
_{0}\right) \}^{-1}S_{\mu i}^{\prime }(y_{i},\mu _{0})$ denoting the
influence function (IF) for $\widehat{\mu }^{\prime }$. The asymptotic
variance of $\widehat{\mu }^{\prime }$ equals the variance of the IF: 
\begin{equation}
\sigma _{sw}^{2}=Var\{ \varphi ^{\prime }\left( y_{i},\mu _{0}\right)
\} =\{B^{\prime }\left( \mu _{0}\right) \}^{-1}Var\{S_{\mu i}^{\prime
}(y_{i},\mu _{0})\}\{B^{\prime }\left( \mu _{0}\right) \}^{-1}.
\label{eqn160}
\end{equation}%
which is the sandwich variance. Since $\varphi ^{\prime }\left( y_{i},\mu
_{0}\right) =y_{i}-\mu _{0}$, $\eqref{eqn160}$ yields $\sigma
_{sw}^{2}=Var\left( y_{i}-\mu _{0}\right) =\mu _{0}\left( 1+\alpha _{0}\mu
_{0}\right) $, the same as the MLE $\widehat{\mu }_{NB}$ for $\mu $ under NB
in (\ref{eqn100}).

Thus even under the mis-specified Poisson, the Poisson score $S_{\mu
i}^{\prime }(y_{i},\mu _{0})$, or influence function $\varphi ^{\prime
}\left( y_{i},\mu _{0}\right) $, remains efficient. Thus, the asymptotic
variance of $\widehat{\mu }^{\prime }$ is still the variance of the IF $%
\varphi ^{\prime }\left( y_{i},\mu _{0}\right) $. However, the asymptotic
variance is given by the sandwich variance $\sigma _{sw}^{2}$, not the
inverse of the Fisher information under Poisson, $I^{^{\prime }}\left( \mu
_{0}\right) =\mu _{0}$. \ 

In the above Example, the NB-based IF and Poisson-based IF for estimating $%
\mu _{0}$ happen to coincide, $\varphi \left( y_{i},\mu _{0}\right) =\varphi
^{\prime }\left( y_{i},\mu _{0}\right) =y_{i}-\mu _{0}$. Thus the
Poisson-based IF\ $\varphi ^{\prime }\left( y_{i},\mu _{0}\right) $ remains
efficient when mis-specified to estimate $\mu $ under NB, i.e., $\varphi
^{\prime }\left( y_{i},\mu _{0}\right) $ remains orthogonal to the nuisance
tangent space (regarding $\alpha $). In fact, $y_{i}-\mu _{0}$ remains
efficient for any other parametric or even non-parametric data generating
process $y_{i}$ (\citealt{tsiatis2006semiparametric}). \ 

For general semiparametric, or restricted moment, models, estimating
functions are generally different when nuisance parameters are
misspecified. However, they remain efficient for estimating their
intended target parameters, so the plug-in rule can be applied to estimate
asymptotic variance of the point estimators defined by the estimating
functions. \ 

\subsubsection*{2. Proof of Variational Independence for Efficient Score}
We defined an efficient score $\boldsymbol{U}_{i}^{eff}\left( \boldsymbol{\theta }\right) $ in Section 3.2. Below we show that by constructing such efficient score, we can address the covariation between $\boldsymbol{U}_{i}\left( \boldsymbol{\theta }\right)$ and $\boldsymbol{V}_{i}^{\top }\left( \boldsymbol{\theta }\right)$, i.e., $\boldsymbol{U}_{i}^{eff}\left( \boldsymbol{\theta }_{0}\right) $ and $\boldsymbol{V}_{i}^{\top }\left( \boldsymbol{\theta }_{0}\right)$ are uncorrelated.
\begin{align*}
E\left \{ \boldsymbol{U}_{i}^{eff}\left( \boldsymbol{\theta }_{0}\right) 
\boldsymbol{V}_{i}^{\top }\left( \boldsymbol{\theta }_{0}\right) \right \} &
=E\left[ \{ \boldsymbol{U}_{i}\left( \boldsymbol{\theta }_{0}\right) -%
\boldsymbol{I}_{12}\boldsymbol{I}_{22}^{-1}\mathbf{V}_{i}\left( \boldsymbol{%
\theta }_{0}\right) \} \boldsymbol{V}_{i}^{\top }\left( \boldsymbol{\theta }%
_{0}\right) \right] \\
& =E\left \{ \boldsymbol{U}_{i}\left( \boldsymbol{\theta }_{0}\right) 
\boldsymbol{V}_{i}^{\top }\left( \boldsymbol{\theta }_{0}\right) \right \} -%
\boldsymbol{I}_{12}\boldsymbol{I}_{22}^{-1}E\left \{ \boldsymbol{V}%
_{i}\left( \boldsymbol{\theta }_{0}\right) \boldsymbol{V}_{i}^{\top }\left( 
\boldsymbol{\theta }_{0}\right) \right \} \\
& =\boldsymbol{I}_{12}\left( \boldsymbol{\theta }_{0}\right) -\boldsymbol{I}%
_{12}\left( \boldsymbol{\theta }_{0}\right) \boldsymbol{I}_{22}^{-1}\left( 
\boldsymbol{\theta }_{0}\right) I_{22}\left( \boldsymbol{\theta }_{0}\right)
\\
& =\mathbf{0}
\end{align*}%

\subsection*{Appendix B. Proof of Theorem 1}
For each cross-fitted sample, by (14) in the main text and $%
\lim_{n\rightarrow \infty }\frac{n_{1}}{n}=\frac{1}{2}$, we \ have:\ 
\begin{equation}
\sqrt{n_{1}}\left( \widehat{\boldsymbol{\mu} }_{1}- \boldsymbol{\mu} _{0}\right) \rightarrow
_{p}N\left( \boldsymbol{0},\Sigma _{\boldsymbol{\mu} }\right) ,\mathbf{\quad }\Sigma _{\boldsymbol{\mu} }=Var\{
\boldsymbol{U}_{i}\left( \boldsymbol{\theta }_{0}^{\prime }\right) \}
\label{eqn580.20}
\end{equation}%

In Theorem 1, we considered the average between two cross-fitted samples as our final point and variance estimator. We provide a proof of Theorem 1 on the asymptotic properties of such estimators below. \ 

Since $E(\boldsymbol{W}_{ui}\left( \boldsymbol{\theta }^{\prime }\right)) = 0$ and $E(\boldsymbol{W}_{vi}\left( \boldsymbol{\theta }^{\prime }\right)) = 0$, we have $\widehat{\boldsymbol{\mu} }_{1} \rightarrow \boldsymbol{\mu}$ and $\widehat{\boldsymbol{\mu} }_{2} \rightarrow \boldsymbol{\mu}$. Thus, $\widehat{\boldsymbol{\mu} }^{sscf} \rightarrow \boldsymbol{\mu}$.

By (13) in main text, the IF\ for $\widehat{\boldsymbol{ \theta }_{k}}^{\prime
}=( \widehat{\boldsymbol{\mu} }_{k},\widehat{\boldsymbol{\psi }}_{k}^{\prime
 \top }) ^{\top }$ has the form $\boldsymbol{\varphi }%
_{ki}^{\prime }\left( \boldsymbol{\theta }_{0}^{\prime }\right) = \boldsymbol{I}\left( 
\boldsymbol{\theta }_{0}^{\prime }\right) \boldsymbol{W}_{ri}$ $\left( k=1,2\right) $.
Since the two samples of the mixture distribution are complementary to each
other, the asymptotic covariance $\Sigma _{\theta ,12}$ between $\widehat{%
\boldsymbol{\theta }_{1}}^{\prime }$ and $\widehat{\boldsymbol{\theta }_{2}}%
^{\prime }$ is the sum of the following terms: 
\begin{eqnarray*}
E\{ \boldsymbol{I}^{-2}\left( \boldsymbol{\theta }_{0}^{\prime }\right) \boldsymbol{W}_{ri}\boldsymbol{W}_{sj}%
\} &=&\left\{ 
\begin{array}{ll}
\boldsymbol{I}^{-2}\left( \boldsymbol{\theta }_{0}^{\prime }\right) E\left(
\boldsymbol{W}_{ri}\boldsymbol{W}_{rj}^{\top }\right)  & \text{for }i\neq j \\ 
\boldsymbol{I}^{-2}\left( \boldsymbol{\theta }_{0}^{\prime }\right) E\left(
\boldsymbol{W}_{ri}\boldsymbol{W}_{si}^{\top }\right)  & \text{for }i=j,r\neq s%
\end{array}%
\right. \text{ } \\
1 &\leq &i,j\leq n,\quad r=u,v,\quad s=u,v
\end{eqnarray*}%
Since each sample is an i.i.d. sequence of $\boldsymbol{W}_{ri}$, $E\left(
\boldsymbol{W}_{ri}\boldsymbol{W}_{rj}^{\top }\right) =0$ for $i\neq j$. \ For $i=j,r\neq s$, we
have:\ 
\begin{align*}
E\left( \boldsymbol{W}_{ri}\boldsymbol{W}_{si}^{\top }\right) & =\left\{ 
\begin{array}{ll}
E\left\{ \left( 
\begin{array}{c}
\boldsymbol{U}_{i} \\ 
\boldsymbol{0}%
\end{array}%
\right) \left( 
\begin{array}{c}
\boldsymbol{0} \\ 
\boldsymbol{V}_{i}%
\end{array}%
\right) ^{\top }\right\} & \text{if }r=u,s=v \\ 
E\left\{ \left( 
\begin{array}{c}
\boldsymbol{0} \\ 
\boldsymbol{V}_{i}%
\end{array}%
\right) \left( 
\begin{array}{c}
\boldsymbol{U}_{i} \\ 
\boldsymbol{0}%
\end{array}%
\right) ^{\top }\right\}  & \text{if }r=v,s=u%
\end{array}%
\right.  \\
& =\left\{ 
\begin{array}{ll}
\left( 
\begin{array}{cc}
\boldsymbol{0} & E\left( \boldsymbol{U}_{i}\boldsymbol{V}_{i}^{\top }\right)  \\ 
\boldsymbol{0} & \boldsymbol{0}%
\end{array}%
\right)  & \text{if }r=u,s=v \\ 
\left( 
\begin{array}{cc}
\boldsymbol{0} & \boldsymbol{0} \\ 
E\left( \boldsymbol{V}_{i}\boldsymbol{U}_{i}^{\top }\right)  & \boldsymbol{0}%
\end{array}%
\right)  & \text{if }r=v,s=u%
\end{array}%
\right. 
\end{align*}%
Thus the asymptotic covariance is 
\begin{align*}
\Sigma _{\boldsymbol{\theta} ,12}& =\frac{1}{2}\left( 
\begin{array}{cc}
\boldsymbol{0} & E\left( \boldsymbol{U}_{i}\boldsymbol{V}_{i}^{\top }\right)  \\ 
\boldsymbol{0} & \boldsymbol{0}%
\end{array}%
\right) +\frac{1}{2}\left( 
\begin{array}{cc}
\boldsymbol{0} & \boldsymbol{0} \\ 
E\left( \boldsymbol{V}_{i}\boldsymbol{U}_{i}^{\top }\right)  & \boldsymbol{0}%
\end{array}%
\right)  \\
& =\frac{1}{2}\left( 
\begin{array}{cc}
\boldsymbol{0} & E\left( \boldsymbol{U}_{i}\boldsymbol{V}_{i}^{\top }\right)  \\ 
E\left( \boldsymbol{V}_{i}\boldsymbol{U}_{i}^{\top }\right)  & \boldsymbol{0}%
\end{array}%
\right) 
\end{align*}%
It follows that the asymptotic covariance between $\widehat{\boldsymbol{\mu} }_{1}$ and $%
\widehat{\boldsymbol{\mu} }_{2}$ is $\Sigma _{\boldsymbol{\mu} ,12}=0$. \ 

Thus by (\ref{eqn580.20}), we have: 
\begin{align*}
\sqrt{n}\left( \widehat{\boldsymbol{\mu}}^{sscf}- \boldsymbol{\mu}_{0}\right) & =\sqrt{n}\{ \frac{1}{2}%
\left( \widehat{\boldsymbol{\mu}}_{1}+\widehat{\boldsymbol{\mu}}_{2}\right) -\boldsymbol{\mu}_{0}\} \\
& =\frac{1}{2}\sqrt{n}\{\left( \widehat{\boldsymbol{\mu}}_{1}-\boldsymbol{\mu}_{0}\right) +\left( 
\widehat{\boldsymbol{\mu}}_{2}-\boldsymbol{\mu}_{0}\right) \}\\
& \rightarrow_{p}N\left( \boldsymbol{0},\Sigma_{\boldsymbol{\mu}}\right)
\end{align*}
Thus $\widehat{\Sigma}_{\boldsymbol{\mu}}^{sscf}$ is a consistent estimator of the asymptotic
variance of $\widehat{\boldsymbol{\mu}}^{sscf}$. \

\subsection*{Appendix C. Inference for PS and OR Models in the Simulation Study}

As in Section~$\ref{sec2}$, let $U_{i}\left( \boldsymbol{\theta }\right) $
denote the causal DRE estimating function for estimating
the ACE$\ \mu _{0}$, and $ \boldsymbol{V}_{i}\left( \boldsymbol{\psi }%
\right) $ the score from the correctly-specified logistic regression PS and $\boldsymbol{T}_{i}\left( \boldsymbol{\xi }\right) $ the score from the
linear OR. \ The scores for the nuisance parameters are
given by: 
\begin{equation*}
\boldsymbol{V}_{i}(\boldsymbol{\psi })=(x_{i}-\eta _{i}(\boldsymbol{\psi }))\mathbf{z}%
_{i},\quad \boldsymbol{T}_{i}\left( \boldsymbol{\xi }\right) =\frac{1}{\sigma ^{2}}%
(x_{i}-Q_{i}(x_{i};\mathbf{z}_{i},\boldsymbol{\xi }))\mathbf{z}_{i}
\end{equation*}%
Let $U_{i}^{\prime }\left( \boldsymbol{\theta }^{\prime }\right) $ and $%
\boldsymbol{V}_{i}^{\prime }\left( \boldsymbol{\psi }^{\prime }\right) $ denote the
respective $U_{i}\left( \boldsymbol{\theta }\right) $ and $\mathbf{V}_{i}\left( 
\boldsymbol{\psi }\right) $ under the mis-specified $\boldsymbol{\psi }%
^{\prime }$. \ Then $U_{i}^{\prime }\left( \boldsymbol{\theta }^{\prime
}\right) $ is the DRE estimating function $U_{i}\left( \boldsymbol{\theta }\right) $
with $\eta _{i}\left( \boldsymbol{\psi }\right) $ replaced by $\eta
_{i}^{\prime }\left( \boldsymbol{\psi }^{\prime }\right) $, while $%
\boldsymbol{V}_{i}^{\prime }\left( \boldsymbol{\psi }^{\prime }\right) $ is given by:\ 
\begin{equation*}
\boldsymbol{V}_{i}{}^{\prime }(\boldsymbol{\psi }^{\prime })=(x_{i}-\eta _{i}^{\prime
}\left( \boldsymbol{\psi }^{\prime }\right) )\mathbf{z}_{i}
\end{equation*}

Inference about $\boldsymbol{\theta}=\left( \mu,\boldsymbol{\psi}^{\top },%
\boldsymbol{\xi}^{\top}\right) ^{^{\top}}$ and $\boldsymbol{\theta}%
^{\prime}=\left( \mu^{\prime},\left( \boldsymbol{\psi}^{\prime}\right)
^{\top},\boldsymbol{\xi}^{\top}\right) ^{\top}$ is based on the respective
joint estimating equations:%
\begin{align}
\text{ Correctly-specified} & :\sum_{i=1}^{n}\boldsymbol{W}_{i}\left( \boldsymbol{\theta }%
\right) =\left( 
\begin{array}{c}
U_{i}\left( \boldsymbol{\theta}\right) \\ 
\boldsymbol{V}_{i}\left( \boldsymbol{\psi}\right) \\ 
\boldsymbol{T}_{i}\left( \boldsymbol{\xi}\right)%
\end{array}
\right) =\mathbf{0}  \label{eqn600} \\
\text{ Mis-specified} & :\sum_{i=1}^{n}\boldsymbol{W}_{i}^{\prime}\left( \boldsymbol{%
\theta}^{\prime}\right) =\left( 
\begin{array}{c}
U_{i}^{\prime}\left( \boldsymbol{\theta}^{\prime}\right) \\ 
\boldsymbol{V}_{i}^{\prime}\left( \boldsymbol{\psi}^{\prime}\right) \\ 
\boldsymbol{T}_{i}\left( \boldsymbol{\xi}\right)%
\end{array}
\right) =\mathbf{0}  \notag
\end{align}
Let $\widehat{\boldsymbol{\theta}}=( \widehat{\mu},\widehat {%
\boldsymbol{\psi}}^{\top},\widehat{\boldsymbol{\xi}}^{\top}) ^{\top}$
and $\widehat{\boldsymbol{\theta}}^{\prime}=\ (\widehat{\mu}^{\prime
},\widehat{\boldsymbol{\psi}}^{\prime \top}, \widehat{%
\boldsymbol{\xi}}^{\prime \top}) ^{\top}$ be the solutions
to the correctly- and mis-specified estimating equations in (\ref{eqn600}),
respectively. \ 

Since $U_{i}\left( \boldsymbol{\theta }\right) $ is efficient under $%
\boldsymbol{\psi }$ \ and $\boldsymbol{\xi }$, following the discussion in
Section 3 the asymptotic variance $\sigma _{\mu }^{2}$ of $\widehat{%
\mu }$ is the variance of $U_{i}\left( \boldsymbol{\theta }_{0}\right) $,
i.e., $\sigma _{\mu }^{2}=Var\{ U_{i}\left( \boldsymbol{\theta }%
_{0}\right) \} $. \ Under the mis-specified PS $\boldsymbol{\psi }%
^{\prime }$, $U_{i}^{\prime }\left( \boldsymbol{\theta }^{\prime }\right) $
is not efficient and $Var\{ U_{i}^{\prime }\left( \boldsymbol{\theta }%
_{0}^{\prime }\right) \} $ is no longer the asymptotic variance of $%
\widehat{\mu }$. \ In this case,\ we first estimate the asymptotic variance $%
\Sigma _{\boldsymbol{\theta} }^{\prime }$ of $\widehat{\boldsymbol{\theta }}^{\prime }$
through joint inference. \ Let $\Sigma _{\boldsymbol{\theta} }^{\prime }$ be partitioned
according to the dimension of $\mu _{0}$, $\boldsymbol{\psi }_{0}^{\prime }$
and $\boldsymbol{\xi }_{0}$: 
\begin{equation*}
\Sigma _{\boldsymbol{\theta} }^{\prime }=\left( 
\begin{array}{ccc}
\Sigma _{11}^{\prime }\left( \boldsymbol{\theta }_{0}^{\prime }\right)  & 
\Sigma _{12}^{\prime }\left( \boldsymbol{\theta }_{0}^{\prime }\right)  & 
\Sigma _{13}^{\prime }\left( \boldsymbol{\theta }_{0}^{\prime }\right)  \\ 
\Sigma _{21}^{\prime }\left( \boldsymbol{\theta }_{0}^{\prime }\right) & \Sigma _{22}^{\prime }\left( \boldsymbol{\theta }_{0}^{\prime }\right)  & 
\Sigma _{23}^{\prime }\left( \boldsymbol{\theta }_{0}^{\prime }\right)  \\ 
\Sigma _{31}^{\prime }\left( \boldsymbol{\theta }_{0}^{\prime }\right)& \Sigma _{32}^{\prime }\left( \boldsymbol{\theta }_{0}^{\prime }\right) & \Sigma _{33}^{\prime }\left( \boldsymbol{\theta }_{0}^{\prime }\right) 
\end{array}%
\right) 
\end{equation*}%
Then $\Sigma _{11}^{\prime }\left( \boldsymbol{\theta }_{0}^{\prime }\right) 
$, not $Var\{ U_{i}^{\prime }\left( \boldsymbol{\theta }_{0}^{\prime
}\right) \} $, is the asymptotic variance of $\widehat{\mu }^{\prime }$%
. \ 

\begin{table}[!htbp]
\centering
\footnotesize
\begin{threeparttable}
\caption{Estimated Pearson correlation between influence function $\widehat{U}$ and the score function contributions of propensity score (PS) under correctly specified and misspecified PS models.}
\label{tab:pearson}
\begin{tabular}{ll ll}
\toprule
\multicolumn{2}{c}{PS correct} & \multicolumn{2}{c}{PS mis-specified} \\
\cmidrule(lr){1-2} \cmidrule(lr){3-4}
Equations & Pearson Correlation & Equations & Pearson Correlation \\
\midrule
$\widehat{U}, V_{0}(\hat{\psi}_{0})$ & $-2.11\times10^{-4}$ & $\widehat{U}, V_{0}(\hat{\psi}_{0}^{\prime})$ & $-5.70\times10^{-3}$ \\
$\widehat{U}, V_{1}(\hat{\psi}_{1})$ & $-5.74\times10^{-4}$ & $\widehat{U}, V_{1}(\hat{\psi}_{1}^{\prime})$ & $-1.64\times10^{-3}$ \\
$\widehat{U}, V_{2}(\hat{\psi}_{2})$ & $-8.60\times10^{-4}$ & & \\
\bottomrule
\end{tabular}
\end{threeparttable}
\end{table}

\begin{table}[!htbp]
\centering
\footnotesize
\begin{threeparttable}
\caption{Summary of propensity score model, and Pearson correlation of the score
function of the propensity score model with the influence function (IF).}
\label{tab:psmodel}
\begin{tabular}{lccc}
\toprule
Covariates & Estimate & P-value & Pearson Correlation with IF \\
\midrule
Intercept       & 0.19  & 0.76        & -0.01 \\
Age             & 0.01  & 0.68        & -0.02 \\
Education       & -2.14 & $<0.001$    & -0.02 \\
Race Hispanic   & -3.16 & $<0.001$    & -0.05 \\
Race White      & -0.92 & $<0.001$    & -0.01 \\
Marital Status  & 0.05  & 0.32        & -0.13 \\
\bottomrule
\end{tabular}
\end{threeparttable}
\end{table}

\begin{table}[!htbp]
\centering
\footnotesize
\begin{threeparttable}
\caption{Comparison of standard errors of the causal estimand, average causal
effect, across different estimators.}
\label{tab:se_comparison}
\begin{tabular}{lc}
\toprule
Estimators & Standard Error \\
\midrule
DRE IF plug-in based without SSCF & 775.8 \\
DRE IF plug-in based with SSCF    & 821.6 \\
Jointly estimated                 & 816.0 \\
\bottomrule
\end{tabular}
\end{threeparttable}
\end{table}

\newpage
\end{document}